\documentclass[twocolumn]{aastex631}

\usepackage{enumerate}
\usepackage{diagbox}
\usepackage{slashbox}
\usepackage{multirow}
\usepackage{amsmath}
\usepackage[utf8]{inputenc}
\usepackage{booktabs}

\let\oldequation\equation
\let\oldendequation\endequation

\renewenvironment{equation}{\linenomathNonumbers\oldequation}{\oldendequation\endlinenomath}
\received{XXX X, XXXX}
\revised{YYY Y, YYYY}
\accepted{ZZZ Z, ZZZZ}

\shorttitle{PAC.\romannumeral5. Quenching of Companion galaxies}
\shortauthors{Zheng et al.}

\begin{document}

\title{PAC. \uppercase\expandafter{\romannumeral5}. The Roles of Mass and Environment in the Quenching of Galaxies}



\correspondingauthor{Y.P. Jing}
\email{ypjing@sjtu.edu.cn}

\author[0000-0001-6575-0142]{Yun Zheng}
\affiliation{Department of Astronomy, School of Physics and Astronomy, Shanghai Jiao Tong University, Shanghai, 200240, People’s Republic of China}

\author[0000-0002-7697-3306]{Kun Xu}
\affil{Department of Astronomy, School of Physics and Astronomy, Shanghai Jiao Tong University, Shanghai, 200240, People’s Republic of China}
\affil{Institute for Computational Cosmology, Department of Physics, Durham University, South Road, Durham DH1 3LE, UK}

\author[0000-0002-4534-3125]{Y.P. Jing}
\affil{Department of Astronomy, School of Physics and Astronomy, Shanghai Jiao Tong University, Shanghai, 200240, People’s Republic of China}
\affil{Tsung-Dao Lee Institute, and Shanghai Key Laboratory for Particle Physics and Cosmology, Shanghai Jiao Tong University, Shanghai, 200240, People’s Republic of China}

\author{Donghai Zhao}
\affil{Key Laboratory for Research in Galaxies and Cosmology, Shanghai Astronomical Observatory, Shanghai, 200030, People’s Republic of China}
\affil{Department of Astronomy, School of Physics and Astronomy, Shanghai Jiao Tong University, Shanghai, 200240, People’s Republic of China}

\author{Hongyu Gao}
\affil{Department of Astronomy, School of Physics and Astronomy, Shanghai Jiao Tong University, Shanghai, 200240, People’s Republic of China}

\author[0000-0002-4574-4551]{Xiaolin Luo}
\affil{Department of Astronomy, School of Physics and Astronomy, Shanghai Jiao Tong University, Shanghai, 200240, People’s Republic of China}

\author[0000-0002-8010-6715]{Jiaxin Han}
\affil{Department of Astronomy, School of Physics and Astronomy, Shanghai Jiao Tong University, Shanghai, 200240, People’s Republic of China}

\author{Yu Yu}
\affil{Department of Astronomy, School of Physics and Astronomy, Shanghai Jiao Tong University, Shanghai, 200240, People’s Republic of China}

\author[0000-0002-1318-4828]{Ming Li}
\affil{National Astronomical Observatories, Chinese Academy of Sciences, Beijing, 100101, People’s Republic of China}

\begin{abstract}

The roles that mass and environment play in the galaxy quenching are still under debate. Leveraging the Photometric objects Around Cosmic webs (PAC) method, we analyze the excess surface distribution $\bar{n}_2w_{\rm{p}}(r_{\rm{p}})$ of photometric galaxies in different color (rest-frame $u-r$) within the stellar mass range of $10^{9.0}M_{\odot}\sim10^{11.0}M_{\odot}$  around spectroscopic massive central galaxies ($10^{10.9}\sim10^{11.7}M_{\odot}$) at the redshift interval $0<z_s<0.7$, utilizing data from the Hyper SuprimeCam Subaru Strategic Program and the spectroscopic samples of Slogan Digital Sky Survey (i.e. Main, LOWZ and CMASS samples). We find that both mass and environment quenching contribute to the evolution of companion galaxies. To isolate the environment effect, we quantify the quenched fraction excess (QFE) of companion galaxies encircling massive central galaxies within $0.01h^{-1}{\rm{Mpc}}<r_{\rm{p}}<20h^{-1}\rm{Mpc}$, representing the surplus quenched fraction relative to the average. We find that the high density halo environment affects the star formation quenching up to about three times of the virial radius, and this effect becomes stronger at lower redshift.  We also find that even after being scaled by the virial radius, the environment quenching efficiency is higher for more massive halos or for companion galaxies of higher stellar mass, though the trends are quite weak. We present a fitting formula that comprehensively captures the QFE across central and companion stellar mass bins, halo-centric distance bins, and redshift bins, offering a valuable tool for constraining galaxy formation models. Furthermore, we have made a quantitative comparison with IllustrisTNG that underscores some important differences, particularly in the excessive quenching of low-mass companion galaxies ($<10^{9.5}M_{\odot}$) in the simulation.
\end{abstract}


\section{Introduction} \label{sec:intro}
Nowadays, it is generally adopted that galaxies can be divided into two main populations: star-forming and passive. The former kind of galaxies are typically young, forming new stars, and have blue colors along with late-type morphologies. The latter kind of galaxies are typically old, do not show star formation activity, and have red colors and early-type morphologies \citep{2003ApJ...594..186B,2004ApJ...600..681B,2003MNRAS.341...54K,2004MNRAS.353..713K,2008A&A...483L..39C,2012MNRAS.424..232W,2014ApJ...788...28V,2019MNRAS.483.5444D,2019MNRAS.488..847P}. In order to understand these galaxy properties, a series of physical processes in galaxy formation and evolution should be taken into account. Among them, galaxy quenching plays a remarkable role in shaping galaxy properties such as color and morphology \citep{2003ApJ...594..186B,2004ApJ...600..681B,2004MNRAS.351.1151B,2008A&A...483L..39C,2013ApJ...777...18M,2019MNRAS.483.5444D,2019MNRAS.488..847P}. Thus, a deep investigation of the galaxy quenching will ultimately be a major step forward to our understanding of galaxy formation and evolution.

Mass quenching and environment quenching are the two dominant modes \citep{2010ApJ...721..193P,2010MNRAS.402.1942C,2011MNRAS.411..675S,2012ApJ...757....4P,2013ApJ...777...18M,2016ApJ...825..113D,2016MNRAS.457.4360Z,2017MNRAS.464..121S,2019MNRAS.483.2851S,2020ApJ...889..156C,2020ApJ...890....7C,2022A&A...668A..69E,2023arXiv231210222T}. Mass quenching refers to all the internal processes that quench the star formation and are expected to be primarily dependent on the galaxy mass. Relying on the characteristic stellar mass regimes, distinct processes have been proposed. In the low stellar mass regime ($\log [M_{\star}/M_{\odot}]<10$), gas outflows, which are attributed to stellar feedback processes such as stellar winds or supernova explosions, have a significant impact on inhibiting star formation \citep{1974MNRAS.169..229L,1986ApJ...303...39D,2008MNRAS.387.1431D}. In the high stellar mass regime ($\log [M_{\star}/M_{\odot}] >10$), active galactic nucleus (AGN) feedback is expected to be more effective in quenching star formation \citep{2006MNRAS.365...11C,2007ApJ...660L..11N,2012ARA&A..50..455F,2013ApJ...776...63F,2014A&A...562A..21C,2018MNRAS.476...12B}. Environment quenching is driven by the interactions between galaxies and their local environment, such as ram pressure stripping \citep{1972ApJ...176....1G,1999MNRAS.304..465M,2017MNRAS.466.1275B,2017ApJ...844...48P,2018ApJ...857...71B,2019ApJ...873...52O,2021PASA...38...35C}, strangulation or starvation \citep{1980ApJ...237..692L,1999MNRAS.304..465M,2011ApJ...732...17N,2015Natur.521..192P}, and harassment \citep{1981ApJ...243...32F,1996Natur.379..613M}. Ram pressure stripping occurs when a galaxy moves through the intracluster medium (ICM). If the ram pressure is strong enough, it may strip a galaxy of its entire cold gas reservoir and give rise to an abrupt quenching of its star formation. Strangulation, which assumes that a galaxy’s gas reservoir is cut off when it is accreted into a larger host system, results in a decline of the galaxy’s star-formation rate as it runs out of the fuel. Harassment refers to the cumulative effect of multiple high-speed impulsive encounters, which can result in the removal of gas.

In the past few years, many studies have focused on which mode dominates the quenching of galaxies. It is generally thought that central galaxies, which are the most massive galaxies residing at the centers of dark matter halos, strongly depend on mass quenching \citep[e.g.,][]{2010ApJ...721..193P}. However, the quenching mechanism for satellite galaxies remains a subject of ongoing debate. A bunch of observational results have shown that satellite galaxies are more likely to be quenched in denser environment \citep{2000ApJ...540..113B,2007ApJ...664..791B,2014ApJ...789..164T,2017MNRAS.464..121S}. While others have shown no or little dependence on the proxies for environment, such as halo mass and cluster-centric distance  \citep{2012ApJ...746..188M,2016ApJ...825..113D,2018MNRAS.475..523L}. To elucidate the diverse influences on galaxy quenching, comparisons between observational results and simulations are invaluable. For example, \cite{2020ApJ...889..156C} explored the roles of environment and stellar mass in quenching galaxies using the merger tree of an $N$-body simulation presented in \cite{2012MNRAS.422..804K}. Additionally, \cite{2022arXiv221100010E} quantitatively examined quenched fractions, gas content, and stellar mass assembly of satellite galaxies around MW/M31-like hosts in TNG50, the highest-resolution run of the IllustrisTNG simulations \citep{2019ComAC...6....2N}. In another study, \cite{2022arXiv220813805P} compared simulated satellite galaxies from the Auriga simulations \citep{2017MNRAS.467..179G} with observed satellite galaxies from the ELVES Survey \citep{2022ApJ...933...47C}, revealing consistent properties such as quenched fraction and the luminosity function. Despite notable advancements in hydrodynamic simulations, satellite properties still hinge on sub-grid physical recipes. An accurate delineation of mass and environment quenching mechanisms holds the potential to provide crucial insights for refining sub-grid physics in galaxy formation models.

\defcitealias{2022ApJ...925...31X}{Paper \uppercase\expandafter{\romannumeral1}}
\defcitealias{2022ApJ...939..104X}{Paper \uppercase\expandafter{\romannumeral3}}
\defcitealias{2023ApJ...944..200X}{Paper \uppercase\expandafter{\romannumeral4}}

To refine galaxy quenching models, it is essential to gather observed galaxy color distributions across broad stellar mass ranges and diverse environments. In this study, we employ the Photometric objects Around Cosmic webs (PAC) method, developed in the inaugural paper of this series \citep{2022ApJ...925...31X} (hereafter \citetalias{2022ApJ...925...31X}) built on the work by \citet{2011ApJ...734...88W}. Leveraging large and deep photometric surveys, we quantify the excess surface density distribution of both blue and red companion galaxies encircling massive central galaxies. Our analysis extends to the very faint end ($10^{9.0}M_{\odot}$) and encompasses redshifts up to $z_s\approx0.7$\footnote{Throughout the paper, we use $z_s$ for spectroscopic redshift, $z$ for the $z$-band magnitude.}. By amalgamating the aforementioned outcomes, we delve into the study of the quenched fraction excess (QFE) of companion galaxies attributed to environmental effects. Our findings reveal that environmental quenching predominantly occurs in high-density regions within massive dark matter halos, up to three times the virial radius ($3r_{\rm {vir}}$). Using the QFE, we quantify the efficiency of quenching in different environments. A fitting model for the QFE is constructed, accounting for dependencies on halo-centric distance, halo mass, companion stellar mass, and redshift. Additionally, we compare our results with those from the TNG300 cosmological hydrodynamical simulation, revealing an over-quenching of low-mass galaxies by environmental factors in TNG.  


We adopt the Planck 2018 $\Lambda$CDM cosmological model \citep{2020A&A...641A...6P} with $\Omega_{\mathrm{m},0} = 0.3111$, $\Omega_{\Lambda,0} = 0.6889$ and $H_0 = 67.66 \,\mathrm{km\,s^{-1}\,Mpc^{-1}}$ throughout the paper.

\section{Data and Method} \label{sec:Data and Method}
\subsection{Photometric objects Around Cosmic webs (PAC)}

We introduced a method to estimate the projected density distribution $\bar{n}_2w_{\rm{p}}(r_{\rm{p}})$ of photometric objects with specific physical properties (e.g., luminosity, mass, color) around spectroscopic objects in \citetalias{2022ApJ...925...31X}:
\begin{equation}
    \bar{n}_2w_{\rm{p}}(r_{\rm{p}}) = \frac{\bar{S}_2}{r_1^2}\omega_{12,\rm{weight}}(\vartheta)\,\,.\label{eq:1}
\end{equation}
Here, $w_{\rm{p}}(r_{\rm{p}})$ and $\omega_{12,\rm{weight}}(\vartheta)$ represent the projected cross-correlation function (PCCF) and the weighted angular cross-correlation function (ACCF) between a specified set of spectroscopically identified galaxies and a sample of photometric galaxies. Additionally, $\bar{n}_2$ and $\bar{S}_2$ denote the mean number density and mean angular surface density of the photometric galaxies, while $r_1$ signifies the comoving distance to the spectroscopic galaxies. The weight in $\omega_{12,\rm{weight}}(\vartheta)$ is incorporated to account for the variation in $r_{\rm{p}}$ at a given angle $\theta$ due to galaxies being at different redshifts, where \( r_{\rm{p}} = r_1\theta \). We employ the Landy–Szalay estimator for the two-point correlation function \citep{1993ApJ...412...64L}. A notable strength of PAC lies in its ability to statistically estimate the rest-frame physical properties for photometric objects. The key steps of PAC are as follows:

\begin{enumerate}[(1)]
    \item Split the spectroscopic catalogs into narrower redshift bins to limit the range of $r_1$ in each bin.
    \item Assume that all photometric objects share the same redshift as the mean redshift in each redshift bin and compute the physical properties of the photometric objects. Consequently, there exists a physical property catalog for the photometric sample in each redshift bin.
    \item Select photometric objects with specific physical properties from the catalog and calculate $\bar{n}_2w_{\rm{p}}(r_{\rm{p}})$ using Equation \ref{eq:1} in each redshift bin. Because foreground and background sources along the line-of-sight (LOS) to a spectroscopic galaxy are distributed statistically in the same way as those along a random LOS, the foreground and background objects with wrong properties can be cancelled out through ACCF, and only photometric objects around the spectroscopic galaxy with the correct redshift and physical properties can contribute to the ACCF.
    \item Aggregate the results from different redshift bins and average them with appropriate weights.
\end{enumerate}

For a more comprehensive understanding, we refer the reader to \cite{2011ApJ...734...88W} and \citetalias{2022ApJ...925...31X}. To validate this method, we compare the $\bar{n}_2w_{\rm{p}}(r_{\rm{p}})$ results obtained by PAC with those directly obtained from the spectroscopic sample. The comparison confirms the reliability of PAC. The details are provided in APPENDIX \ref{nwpcheck}.

\subsection{Observational data}
In this subsection, we outline the data from spectroscopic and photometric catalogs utilized in this paper across three distinct redshift ranges.

\subsubsection{Selection of central galaxies}
Focusing on central galaxies within the spectroscopic sample, we define centrals as galaxies which do not have more massive neighbors within a distance of $30r^{\rm nb}_{\rm{vir}}$ along the line-of-sight and within $3r^{\rm nb}_{\rm{vir}}$ perpendicular to the line-of-sight. Here, $r^{\rm nb}_{\rm{vir}}$ denotes the virial radius of the more massive galaxies in comparison. We adopt this relatively conservative criterion for central galaxy selection, recognizing that the environmental impact of a halo extends up to approximately $3r_{\rm vir}$ as we will show below.  As shown by \citet{2014MNRAS.440..762O},  about thirteen percent of the central galaxies thus defined may not be at the centers of the host halo gravitational wells. Since most of these off-centering galaxies are still with the gravitational well and the fraction is small, we believe this population should not have a significant impact on our results below.

\begin{table*}
\caption{The marginalized posterior PDFs of the parameters. $M_0$ is in unit of $h^{-1}M_{\odot}$ and $k$ is in unit of $M_{\odot}$.}\label{DP model}
\begin{tabular}{ccccccc}\\
\hline \hline  redshift  &  model  & ${\rm{log}_{10}}(M_0)$ & $\alpha$ & $\beta$ & ${\rm{log}}_{10}(k)$ & $\sigma$ \\
\hline 
$z_s<0.2$ & \text { DP } & $11.800_{-0.021}^{+0.021}$ & $0.299_{-0.012}^{+0.011}$ & $1.838_{-0.018}^{+0.018}$ & $10.331_{-0.019}^{+0.020}$ & $0.214_{-0.008}^{+0.008}$ \\

$0.2<z_s<0.4$ & \text { DP } & $11.641_{-0.012}^{+0.013}$ & $0.433_{-0.006}^{+0.006}$ & $2.119_{-0.021}^{+0.021}$ & $10.121_{-0.011}^{+0.011}$ & $0.187_{-0.003}^{+0.003}$ \\

$0.5<z_s<0.7$ & \text { DP } & $11.681_{-0.011}^{+0.011}$ & $0.438_{-0.007}^{+0.008}$ & $2.531_{-0.035}^{+0.037}$ & $10.134_{-0.011}^{+0.011}$ & $0.212_{-0.003}^{+0.003}$ \\
\hline
\end{tabular}
\end{table*}

In this analysis, we determine the values of $r_{\rm{vir}}$ by replicating the methodology outlined in \cite{2023ApJ...944..200X} (hereafter \citetalias{2023ApJ...944..200X}), leveraging the high-resolution {\texttt{Jiutian}} N-body simulation \citep{jiutian}. Our objective is to constrain the stellar-halo mass relations (SHMR) under the Planck 2018 cosmology through the application of the subhalo abundance matching (SHAM) method. The {\texttt{Jiutian}} simulation features $6144^3$ dark matter particles within a periodic box of $1000 h^{-1}\rm{Mpc}$, providing a sufficiently large volume for robustly modeling $\bar{n}_2w_{\rm{p}}$ up to scales of 20 $h^{-1}$Mpc. Dark matter halos are identified using the Friends-of-Friends method, and subhalos are traced using {\texttt{HBT+}} \citep{2012MNRAS.427.2437H,2018MNRAS.474..604H}. We adopt the double power law form (DP model in \citetalias{2023ApJ...944..200X}) for the SHMR:

\begin{equation}
M_*=\left[\frac{2}{\left(\frac{M_{\rm acc}}{M_0}\right)^{-\alpha}+\left(\frac{M_{\rm acc}}{M_0}\right)^{-\beta}}\right] k
\end{equation}

\begin{figure*}[htb!]
\begin{center}
\includegraphics[width=1.08\textwidth]{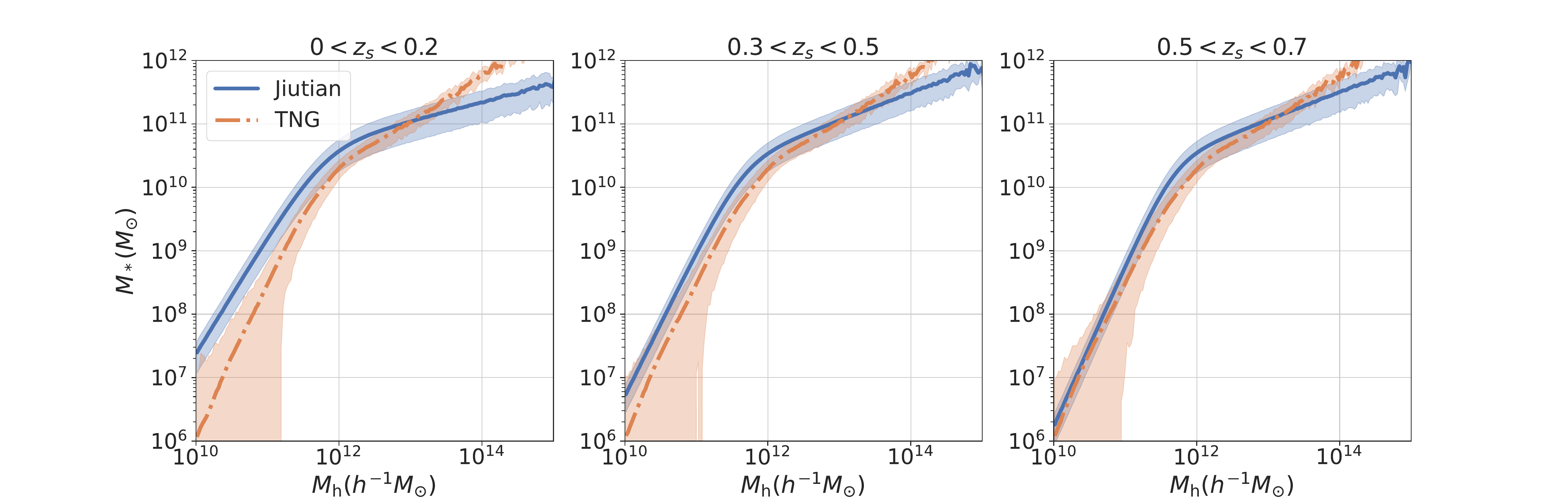}
\end{center}
\caption{The central stellar-halo mass relation of Jiutian and TNG for three distinct redshift bins: $0<z_s<0.2$(left), $0.3<z_s<0.5$(medium), and $0.5<z_s<0.7$(right). Solid blue lines are for Jiutian simulation, dotted red lines are for TNG.}
\label{SHMR}
\end{figure*}

Here $M_*$ is the stellar mass, $M_{\rm{acc}}$ is defined as the viral mass $M_{\rm{vir}}$ of the halo at the time when the galaxy was last the central dominant object. The scatter in $log(M_*)$ at a given $M_{\rm{acc}}$ is described with a Gaussian function of the width $\sigma$. $\alpha$ and $\beta$ represent the slopes of the SHMR at high and low mass ends respectively.
The constrained parameters for the DP model across three redshift ranges are detailed in Table \ref{DP model}. For the purposes of this paper, we employ interpolation to obtain parameter values for the intermediate redshift bin $0.3<z_s<0.5$. The SHMR for different redshift bins in this work are shown in Figure \ref{SHMR}. To derive the halo virial mass $M_{\rm{vir}}$, we use the stellar mass of the central galaxy based on the SHMR. Subsequently, we calculate the corresponding $r_{\rm{vir}}$ by computing mean values within each stellar mass bin. The values of $r_{\rm{vir}}$ and the standard deviation for various central stellar mass bins utilized in this study are presented in Table \ref{rvir_jiutian_TNG}. The values of $M_{\rm{vir}}$ and the scatter for various central stellar mass bins are presented in Table \ref{mvir_jiutian}.
In this work, we use the mean values to construct our model. 

\begin{table*}
\caption{$r_{\rm{vir}}$ derived from Jiutian and TNG for different central stellar mass bins at three different redshift bins, with $r_{\rm{vir}}$ in unit of $h^{-1}\rm{Mpc}$.\label{rvir_jiutian_TNG}}
\centering
\begin{tabular}{cccccc}\\
\hline \hline
\multirow{2}{*}{simulation}&\multirow{2}{*}{redshift bin} & \multicolumn{4}{c}{mass bin [${\rm{log}_{10}}(M_*/M_{\odot})$]} \\
\cline{3-6}
      & & $10.9-11.1 $ & $11.1-11.3 $& $11.3-11.5 $ & $11.5-11.7 $ \\
\hline
    \multirow{3}{*}{Jiutian}&  $z_s<0.2$  & $0.355\pm 0.131$ & $0.455\pm 0.186$ & $0.597 \pm 0.254$ & $0.782\pm 0.323$\\

    & $0.3<z_s<0.5$ &  & &$0.655\pm 0.211$ & $0.848 \pm 0.261$\\

    & $0.5<z_s<0.7$ &  & & $0.622 \pm 0.207$ & $0.795 \pm 0.256$ \\
    
\hline
    \multirow{3}{*}{TNG} & $z<_s0.2$ & $0.417\pm{0.069}$ & $0.515\pm{0.082} $ & $0.644\pm{0.094}$ & $0.789\pm{0.104}$\\

    & $0.3<z_s<0.5$  & & &$0.681\pm{0.096}$ & $0.846\pm{0.117}$\\

    & $0.5<z_s<0.7$  & & & $0.694\pm{0.102}$ & $0.848\pm{0.117}$ \\
    
\hline    
\end{tabular}
\end{table*}

\begin{table*}
\caption{$\rm{log_{10}}(M_{\rm{vir}})$ derived from Jiutian for different central stellar mass bins at three different redshift bins, with $M_{\rm{vir}}$ in unit of $h^{-1}M_{\odot}$.\label{mvir_jiutian}}
\centering
\begin{tabular}{cccccc}\\
\hline \hline
\multirow{2}{*}{simulation}&\multirow{2}{*}{redshift bin} & \multicolumn{4}{c}{mass bin [${\rm{log}_{10}}(M_*/M_{\odot})$]} \\
\cline{3-6}
      & & $10.9-11.1 $ & $11.1-11.3 $& $11.3-11.5 $ & $11.5-11.7 $ \\
\hline
    \multirow{3}{*}{Jiutian}&  $z_s<0.2$  & ${12.859\pm{0.265}}$ & ${13.217\pm{0.263}}$ & ${13.575\pm{0.290}}$ & ${13.917\pm{0.352}}$\\

    & $0.3<z_s<0.5$ &  & &${13.519\pm{0.590}}$ & ${13.844\pm{0.791}}$\\

    & $0.5<z_s<0.7$ &  & & ${13.424\pm{0.523}}$ & ${13.739\pm{0.637}}$ \\
\hline      
\end{tabular}
\end{table*}

\subsubsection{Redshift range: $0.0\sim 0.2$}\label{redshift1}
For the photometric sample, we use the {\tt {datasweep}} catalog of the SDSS DR13 \citep{2017ApJS..233...25A}. It contains a subset of the data from the full SDSS photometric catalogs, which is sufficient for our study. We select all galaxies with $r$ band apparent Model magnitudes $ r < 21.0$. We also check the bitmask called {\tt {RESOLVE\_STATUES}} to select unique objects in the catalog. Then we trim the catalog by the MANGLE software (\citet{2004MNRAS.349..115H};\citet{2008MNRAS.387.1391S}) according to the geometry with the bright star masks cut out provided by the New York University Value Added Galaxy Catalog(NYU-VAGC)\footnote{http://sdss.physics.nyu.edu/vagc/} \citep{2005AJ....129.2562B}. We also restrict the galaxies to the continuous area of the north Galactic cap(NGC). Finally, we obtain a photometric catalog with five bands $ugriz$. We also construct a random catalog with the same selection criteria as the photometric sample by using the MANGLE software.

We employ the spectral energy distribution (SED) code CIGALE \citep{2019A&A...622A.103B} to compute the physical properties (e.g., stellar mass and rest-frame colors) of galaxies. The stellar population synthesis models of \citet{2003MNRAS.344.1000B} and the initial mass function from \cite{2003PASP..115..763C} are selected. We assume a delayed star formation history $\phi(t) \approx t \exp(-t/\tau)$, where $\tau$ spans from $10^7$ yr to $1.258 \times 10^{10}$ yr with an equal logarithmic interval $\Delta \log\tau = 0.1$. The extinction law of \cite{2000ApJ...533..682C} is adopted, with dust reddening in the range $0<E(B-V)<0.5$. Moreover, we set three metallicities, $Z / Z_\odot$ equal to 0.4,1 and 2.5, respectively, where $Z_\odot$ is the metallicity of the Sun. 

The spectroscopic galaxy sample is derived from the SDSS Data Release 7 \citep[DR7,][]{2009ApJS..182..543A} Main sample, which is encompassed in the NYU-VAGC. Initially, we select galaxies within the continuous region of the NGC with redshifts ranging from $0.0 < z_s < 0.2$. The physical properties of these galaxies are determined through SED modeling with the five bands $ugriz$. By applying the previously mentioned central selection criterion, we obtain the spectroscopic catalog. A random catalog for this spectroscopic sample is also constructed using MANGLE. The final effective area is 7032 deg$^2$.

\subsubsection{Redshift range: $0.3\sim 0.5$}\label{redshift2}
We utilize the \textbf{Wide} layer photometric catalog from the second public data release (PDR2) of the Hyper SuprimeCam Subaru Strategic Program (HSC-SSP; \citet{2019PASJ...71..114A}) to construct the photometric sample within this redshift range. We select sources within the footprints observed with all five bands $(grizy)$ to ensure sufficient bands for accurate physical property estimation through SED modeling. The $5\sigma$ depth for $grizy$ are 26.6, 26.2, 26.2, 25.3 and 24.5, respectively. We mask sources around bright objects using the {\tt{\{grizy\}\_mask\_pdr2\_bright\_objectcenter}} flag provided by the HSC collaboration \citep{2018PASJ...70S...7C}. Additionally, the {\tt{\{grizy\}\_extendedness\_value}} flag is employed to exclude stars from the sample.

For the spectroscopic catalog, we utilize the LOWZ sample and the CMASS sample from the Baryon Oscillation Spectroscopic Survey (BOSS; \citet{2012ApJS..203...21A}; \citet{2012AJ....144..144B})\footnote{https://data.sdss.org/sas/dr12/boss/lss/}.The LOWZ database has a smaller footprint than CMASS because the galaxies from the first nine months of the BOSS observation are removed due to the incorrect star-galaxy separation criterion. In total, the LOWZ database covers 8337 $\rm{deg}^2$ of the sky, with 5836 $\rm{deg}^2$ in the NGC and 2501 $\rm{deg}^2$ in the SGC. CMASS covers 9376 $\rm{deg}^2$ of the sky, with 6851 $\rm{deg}^2$ in the NGC and 2525 $\rm{deg}^2$ in the SGC. Initially, we select galaxies in the redshift range $0.3\sim 0.5$. Subsequently, we match this spectroscopic catalog with the HSC photometric sample to obtain magnitudes in the $grizy$ bands for each LOWZ and CMASS galaxy, allowing us to calculate their physical properties. Next, we select central galaxies to obtain the final spectroscopic catalog.

Moreover, we construct a random catalog for the photometric sample using the same selection criteria from the HSC database. Regarding the spectroscopic catalog, we emphasize the need to generate separate random catalogs for the LOWZ sample and the CMASS sample due to their different footprints. We first generate a random catalog in the HSC footprint and then apply the all LSS geometry mask\footnote{https://data.sdss.org/sas/dr12/boss/lss/geometry/} provided by BOSS. For the LOWZ catalog, we apply {\tt{mask\_DR12v5\_LOWZ\_South.ply}} and {\tt{mask\_DR12v5\_LOWZ\_North.ply}} to obtain the corresponding random sample. Similarly, for the CMASS catalog, we use {\tt{mask\_DR12v5\_CMASS\_South.ply}} and {\tt{mask\_DR12v5\_CMASS\_North.ply}}. From these two random samples, we calculate the final effective areas of the LOWZ and CMASS catalogs as 338 $\rm{deg}^2$ and 458 $\rm{deg}^2$, respectively. The final spectroscopic sample and its random sample are the combination of the LOWZ and CMASS corresponding samples.

\subsubsection{Redshift range: $0.5\sim 0.7$}\label{redshift3}
In this redshift range, we employ the same HSC photometric sample mentioned above as the photometric catalog.

For the spectroscopic sample in this redshift range, we select the CMASS sample as mentioned previously. Initially, we choose galaxies within the redshift range $0.5\sim 0.7$. Subsequently, we follow the same steps mentioned earlier to construct both the spectroscopic catalog and the random catalog. 

\subsection{Completeness and designs}
Building upon previous studies \citep{2022ApJ...939..104X} (hereafter \citetalias{2022ApJ...939..104X}), we define the stellar mass range for central galaxies from spectroscopic samples as $[10^{10.9}, 10^{11.7}]M_{\odot}$ for $z_s<0.2$ and $[10^{11.3}, 10^{11.7}]M_{\odot}$ for both $0.3<z_s<0.5$ and $0.5<z_s<0.7$. In the three redshift bins, there are 73446, 4472, and 10039 massive central galaxies, respectively. Subsequently, we divide both the $0.3<z_s<0.5$ and $0.5<z_s<0.7$ samples into two narrower redshift bins with equal bin widths for the PAC measurements. Considering the faster changes in comoving distance at lower redshifts, the $z_s<0.2$ sample is further split into three redshift bins: $[0.05,0.1]$, $[0.1,0.15]$, and $[0.15,0.2]$. Galaxies with $z_s<0.05$ are excluded from the PAC measurements.

Regarding the completeness of the photometric samples, we employ the methodology used in \citetalias{2022ApJ...925...31X} and \citetalias{2022ApJ...939..104X}. The $r$ band (for $z_s<0.2$) and $z$ band (for $0.3<z_s<0.5$ and $0.5<z_s<0.7$) $10\sigma$ PSF depth is adopted as the depth for extended sources to determine the magnitude limit. As demonstrated in \citetalias{2022ApJ...925...31X} (see Figure 1) and \citetalias{2022ApJ...939..104X} (see Figure 3), there is a clear correlation between stellar mass and magnitude in three redshift ranges. Therefore, the magnitude limit can be used to derive a complete sample for a specified stellar mass. Subsequently, we calculate the $r$ band (for $z_s<0.2$) and $z$ band (for $0.3<z_s<0.5$ and $0.5<z_s<0.7$) completeness limit $C_{95}(M_{\star})$, where $95\%$ of the galaxies are brighter than $C_{95}(M_{*})$ in the $r$ or $z$ band for a given stellar mass $M_{*}$. Considering the stellar mass - magnitude relation, we only utilize data in survey footprints deeper than $C_{95}(M_{*})$ for the mass bin of interest. According to Figure 3 of \citetalias{2022ApJ...939..104X}, for the SDSS photometric sample with $r<21.0$, the complete stellar masses are approximately $10^{8.5}M_{\odot}$, $10^{8.8}M_{\odot}$, $10^{9.2}M_{\odot}$, $10^{9.6}M_{\odot}$ at redshifts 0.075, 0.1, 0.15, and 0.2, respectively. Therefore, we calculate $\bar{n}_2w_{\rm{p}}(r_{\rm{p}})$ for the photometric catalog of [$10^{9.2},10^{9.5}]M_{\odot}$ at $z_s<0.15$, and the entire redshift range ($z_s<0.2$) is used for photometric objects more massive than $10^{9.5}M_{\odot}$. For HSC photometric samples at higher redshift ranges, following \citetalias{2022ApJ...925...31X}, we find that $10^{9.0}M_{\odot}$ is complete. The final specifications for the measurements of $\bar{n}_2w_{\rm{p}}(r_{\rm{p}})$ are summarized in Table \ref{sample}.

\begin{table*}
\caption{Final designs for the PAC measurements}\label{sample}
\begin{tabular}{ccccc}
\hline\hline redshift & Survey & $\begin{array}{c}\operatorname{Central}^{\mathrm{a}} \\
\left(M_{\odot}\right)\end{array}$ & $\begin{array}{c}\operatorname{Companion}^{\mathrm{b}} \\
\left(M_{\odot}\right)\end{array}$ & PAC redshift bins \\
\hline$[0.05,0.2]$ & Main & {$\left[10^{10.9}, 10^{11.7}\right]$} & {$\left[10^{9.2}, 10^{9.5}\right],\left[10^{9.5}, 10^{11.0}\right]$} & {$[0.05,0.1],[0.1,0.15],[0.15,0.2]$} \\
\hline$[0.3,0.5]$ & LOWZ $\&$ CMASS & {$\left[10^{11.3}, 10^{11.7}\right]$} & {$\left[10^{9.0}, 10^{11.0}\right]$} & {$[0.3,0.4],[0.4,0.5]$} \\
\hline$[0.5,0.7]$ & CMASS & {$\left[10^{11.3}, 10^{11.7}\right]$} & {$\left[10^{9.0}, 10^{11.0}\right]$} & {$[0.5,0.6],[0.6,0.7]$} \\
\hline
\end{tabular}

${ }^{\mathrm{a}}$ Stellar mass ranges of central spectroscopic sample with an fiducial equal logarithmic bin width of $10^{0.2} M_{\odot}$.

${ }^{\mathrm{b}}$ Stellar mass ranges of companion photometric sample with an fiducial equal logarithmic bin width of $10^{0.5} M_{\odot}$ except the first bin for the SDSS photometric sample.
\end{table*}

\subsection{IllustrisTNG data}
We compare our results in observations to the IllustrisTNG simulations. The IllustrisTNG simulations \citep{2018MNRAS.480.5113M,2018MNRAS.475..624N,2018MNRAS.475..648P,2018MNRAS.475..676S,2019ComAC...6....2N} are a series of cosmological magneto-hydrodynamical simulations that model comprehensive processes related to galaxy formation and evolution. They are built and improved on the original Illustris project \citep{2014Natur.509..177V,2015A&C....13...12N,2015MNRAS.452..575S} and are to-date one of the most advanced versions of large hydrodynamical simulations in a cosmological context.

In this work, we make use of the simulation called TNG300-1, which is carried with the Planck 2015 $\Lambda \rm{CDM}$ cosmological model with the best-fit parameters: matter density $\Omega_{\mathrm{m},0} = 0.3089$, baryonic density $\Omega_{\mathrm{b},0} = 0.0486$, cosmological constant $\Omega_{\Lambda,0} = 0.6911$, Hubble constant $H_0 = 100 h \,\mathrm{km\, s^{-1}\,Mpc^{-1}}$ with $h = 0.6774$, normalization $\sigma_8 = 0.8159 $ and spectral index $n_s = 0.9667$ \citep{2016A&A...594A..13P}. TNG300-1 has a periodic box with a side length of $205 \mathrm{Mpc}\,h^{-1}$ and saves 100 snapshots from $z_s = 127$ to $z_s = 0$. We use the catalog containing synthetic stellar photometry (i.e. colors) provided by \citet{2018MNRAS.475..624N} to study properties for galaxies. We adopt the stellar mass defined as the sum of all stellar particles within twice the stellar half-mass radius, and the viral radius defined as comoving radius of a sphere centered at the {\tt GroupPos} of this group whose mean density is $\Delta_c$  times the critical density of the Universe, at the time the halo is considered. $\Delta_c$ is taken from the solution of the collapse of a spherical top-hat perturbation\citep{1998ApJ...495...80B}.
In keeping compatible with the observational data, we choose three snapshots 67, 72, and 91 corresponding to redshift 0.5, 0.4, and 0.1 respectively\footnote{Because the catalog which contains synthetic stellar photometry does not include the snapshot at $z_s = 0.6$, we choose the snapshot at $z_s = 0.5$ instead.}. The values of $r_{\rm{vir}}$ corresponding to the different central stellar mass bins in TNG are displayed in Table \ref{rvir_jiutian_TNG}.

\subsection{Selection of Quiescent and Star-forming Galaxies}\label{galaxy_catalog}
To obtain the quenched fraction, we categorize galaxies into “red passive” and “blue star-forming” samples, distinguished by a rest-frame $u-r$ color cut. Here we take full use of the available spectroscopic SDSS DR7 Main sample as mentioned above to make the cut. We select objects within the redshift range $0.0 < z < 0.2$ located in the continuous region of NGC. By employing Model magnitudes in the five bands $ugriz$, we extract physical properties (e.g., stellar mass and rest-frame colors) for these galaxies through the SED fitting code CIGALE. It is noted that, based on prior studies \citep[e.g.,][]{2003MNRAS.341...54K,2004MNRAS.353..713K}, the color cut may exhibit a weak dependence on mass and can shift towards bluer colors at higher redshifts.

\begin{figure}[htb!]
\includegraphics[width=0.55\textwidth]{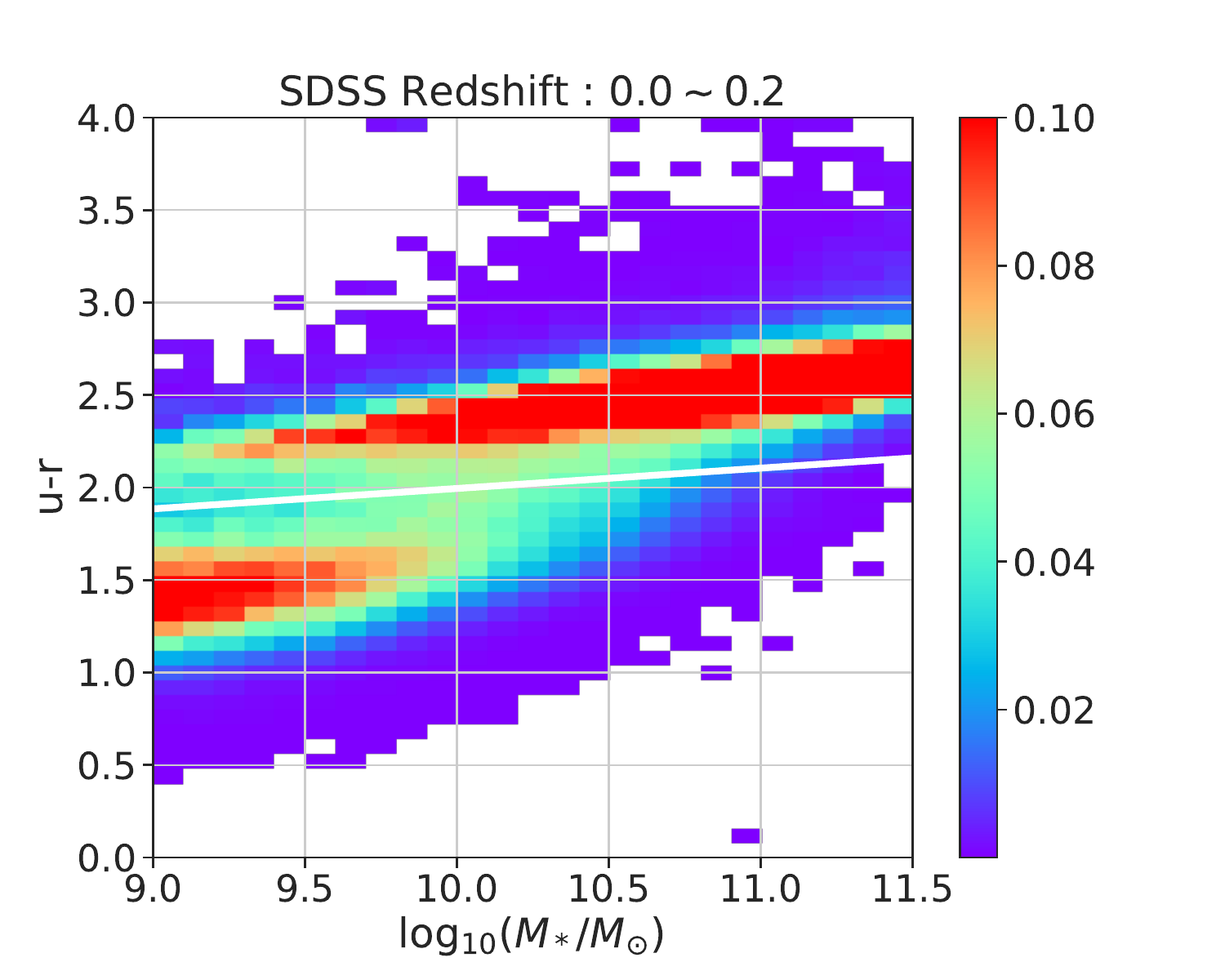}
\caption{Vmax weighted color distribution in SDSS with the dividing lines used to split galaxies into blue and red. The color bar represents the number of galaxies that are normalized by the total number of galaxies in each stellar mass bin.}
\label{redshift_colorcut}
\end{figure}

Figure \ref{redshift_colorcut} shows the color distribution with a $1/V_{\rm{max}}$ weight for each galaxy, where $V_{\rm{max}}$ is the volume corresponding to $z_{\rm{max}}$, the maximum redshift over which the galaxy can pass our sample selection criteria. The dividing lines we adopt in Figure \ref{redshift_colorcut} are given as follows:
\begin{equation}
u - r  = 0.11 \log M_{\star} + 0.895 \label{colorcut_eq}
\end{equation}
We utilize the same color cut across all three redshift bins. The advantage of this criterion, compared to different redshift-dependent color cuts used in literature, is that we use an absolute rule to define the quenching. With this definition, we also find the parameters in the Quenched Fraction Excess (QFE)  model, as will be shown in Section \ref{sec:QFE}, exhibit monotonic redshift dependencies, facilitating the construction of a universal model. Our results can also be easily compared with hydrodynamical simulation with the simple cut criterion.

\section{Results} \label{sec:Results}

\subsection{the excess surface density distribution}
In this section, we estimate the excess surface density distribution $\bar{n}_2w_{\rm{p}}(r_{\rm{p}})$ of different stellar masses and colors for the three redshift ranges according to Table \ref{sample}. Here we adopt the jackknife resampling technique\citep{efron1982jackknife} to evaluate the statistical error. The mean value of $\bar{n}_2w_{\rm{p}}(r_{\rm{p}})$ for galaxies with different stellar mass can be obtained by:

\begin{equation}
\bar{n}_{2} w_{\rm{p}}\left(r_{\rm{p}}\right) = \frac{1}{N_{\mathrm{sub}}} \sum_{k=1}^{N_{\mathrm{sub}}} \bar{n}_{\rm{2}, \rm{k}} w_{\rm{p}, k}\left(r_{\rm{p}}\right)
\end{equation}

The corresponding error is calculated as:

\begin{equation}
\sigma^{2} = \frac{N_{\text{sub}}-1}{N_{\text{sub}}} \sum_{k=1}^{N_{\text{sub}}} \left(\bar{n}_{2, k} w_{\rm{p}, k}\left(r_{\rm{p}}\right) - \bar{n}_{2} w_{\rm{p}}\left(r_{\rm{p}}\right)\right)^{2}
\end{equation}

Here, $N_{\rm{sub}}$ represents the number of jackknife realizations, $\bar{n}_{2, k} w_{\rm{p}, k}\left(r_{\rm{p}}\right)$ denotes the excess of the projected density for the $k$th realization. In this work, we adopt $N_{\rm{sub}}=50$.

To investigate the color distribution of companions, we partition the photometric catalog into red and blue subsamples based on Equation \ref{colorcut_eq}. Following the previously outlined steps, we calculate $\bar{n}_2w_{\rm{p}}(r_{\rm{p}})$ separately for red and blue galaxies. Considering deblending issues in HSC \citep{2021ApJ...919...25W,2021MNRAS.500.3776W}, we adopt an inner radius cut of $0.1 h^{-1}\rm{Mpc}$ for redshift bins $0.3<z_s<0.5$ and $0.5<z_s<0.7$.

\begin{figure*}[h!]
\begin{center}
\includegraphics[width=1.0\textwidth]{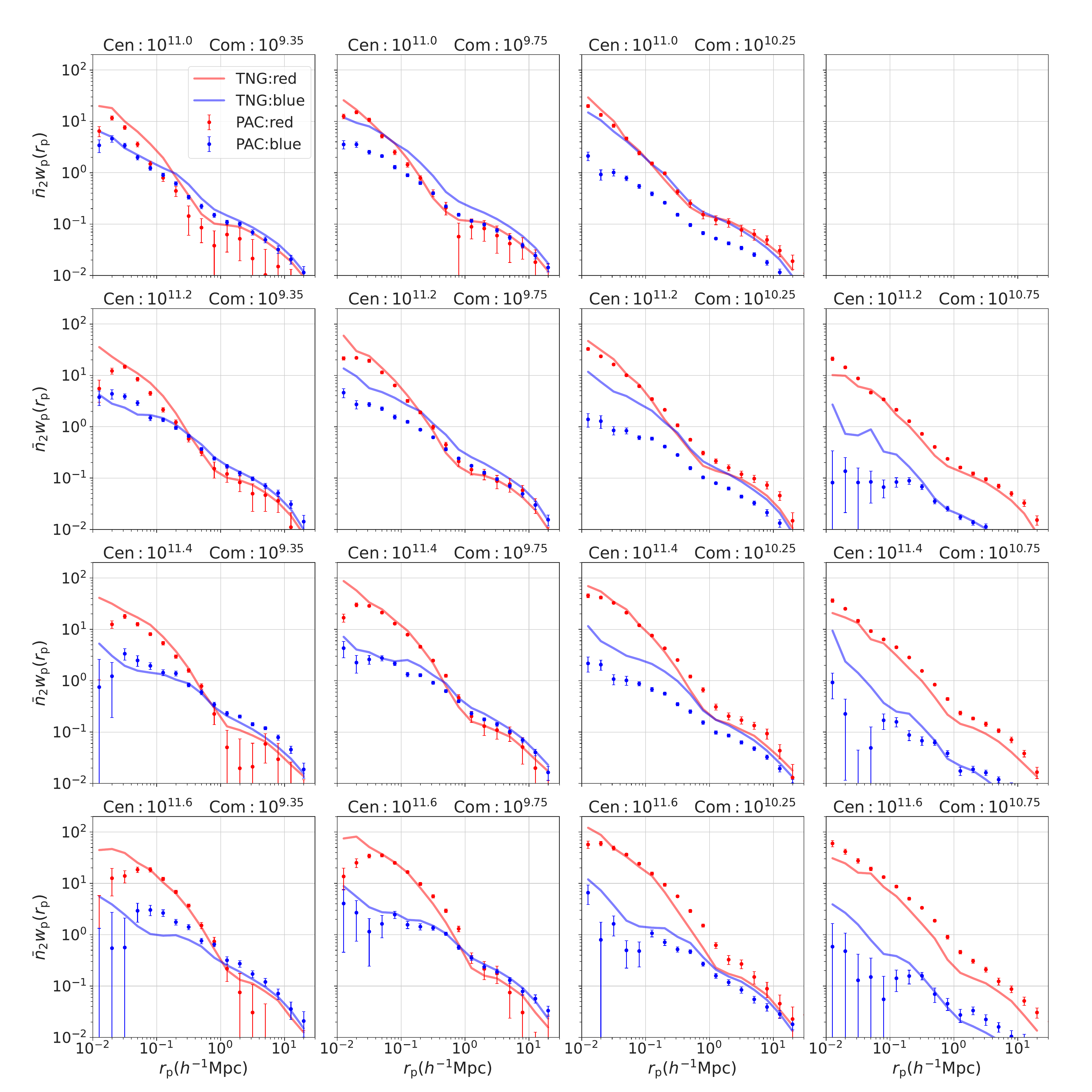}
\end{center}
\caption{Measurements of $\bar{n}_2w_{\rm{p}}(r_{\rm{p}})$ for blue and red subsamples in redshift range $0<z_s<0.2$. The same row represents the results for the same central galaxy mass bin (from top to bottom: $[10^{10.9},10^{11.1},10^{11.3},10^{11.5},10^{11.7}]M_{\odot}$). The four columns from left to right display the results for four distinct mass bins of the photometric sample. The title of each subplot indicates the corresponding central and photometric stellar mass. Colored dots with error bars are the results from observations. Lines are from TNG. Here red color corresponds to red subsamples, blue color corresponds to blue subsamples for both observations and simulations.}
\label{nwp_0_2}
\end{figure*}

\begin{figure*}[htbp]
\begin{center}
\includegraphics[width=1.03\textwidth]{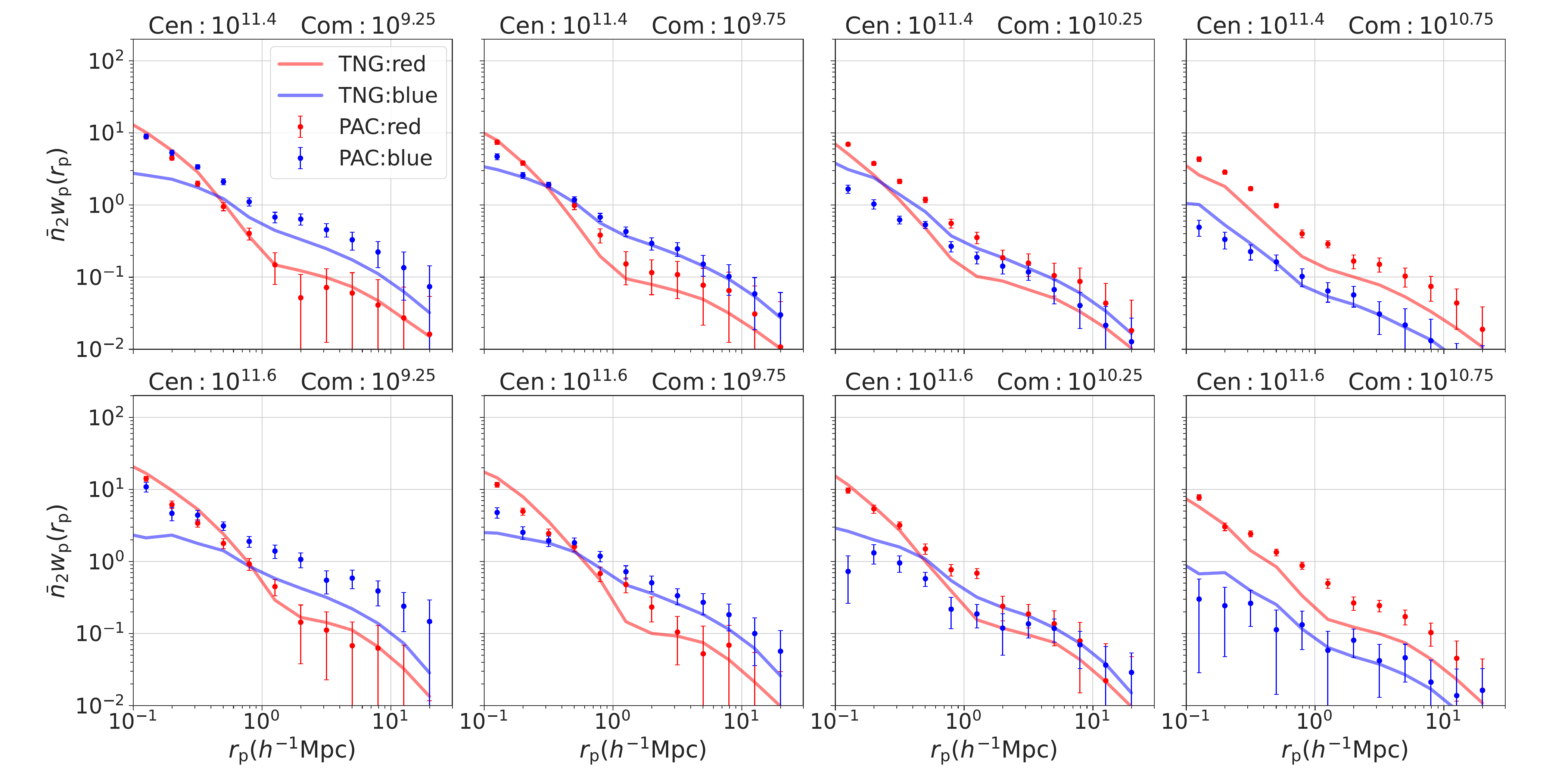}
\end{center}
\caption{The same as Figure \ref{nwp_0_2} but for redshift range $z_s=0.3\sim0.5$. The same row represents the results for the same central galaxy mass bin (from top to bottom: Top:$10^{11.4}M_{\odot}$, Bottom:$10^{11.6}M_{\odot}$).}
\label{nwp_3_5}
\end{figure*}

\begin{figure*}[htbp]
\begin{center}
\includegraphics[width=1.03\textwidth]{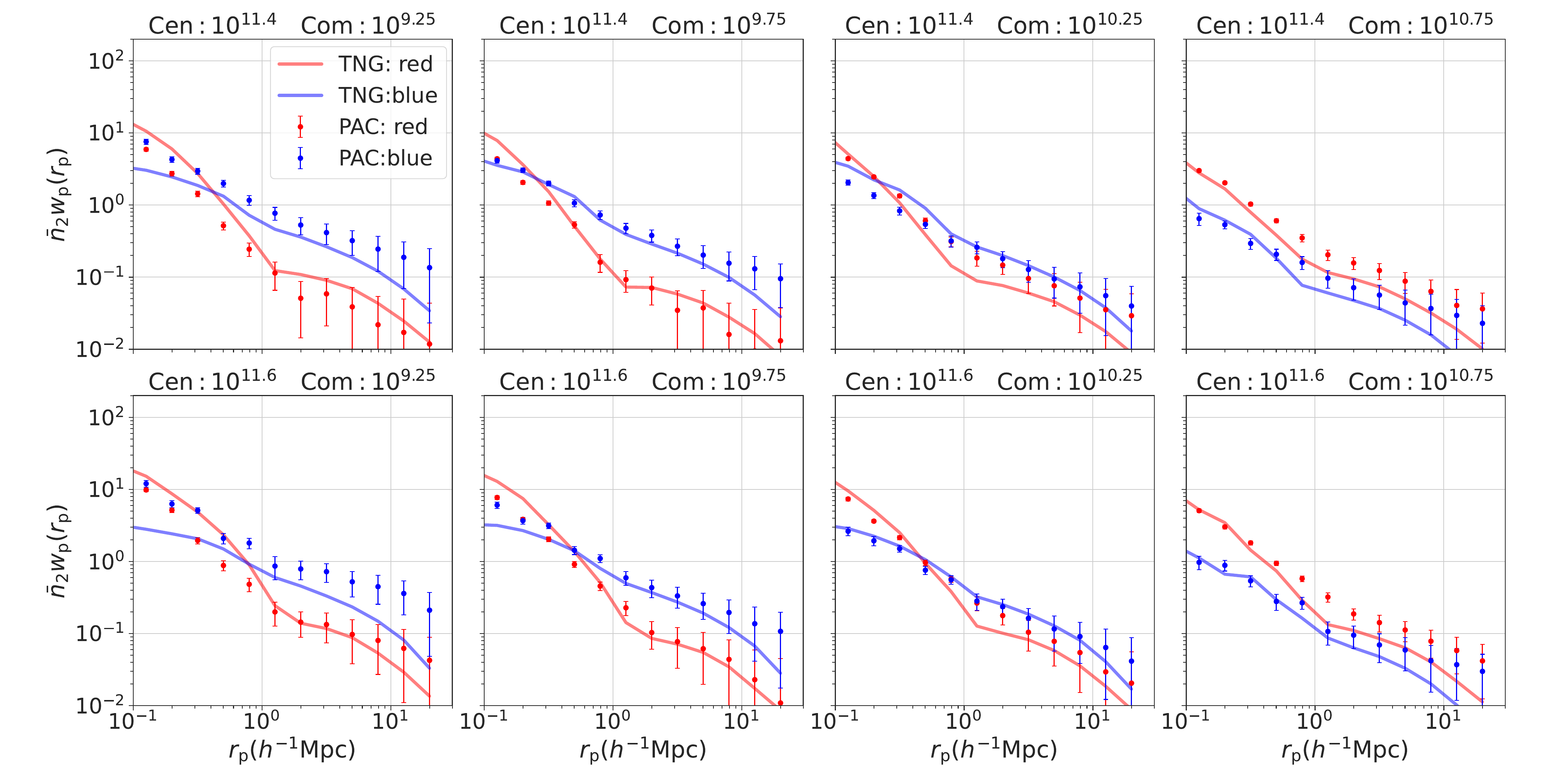}
\end{center}
\caption{The same as Figure \ref{nwp_3_5} but for redshift range $z_s=0.5\sim0.7$.}
\label{nwp_5_7}
\end{figure*}

Additionally, we compute $\bar{n}_2w_{\rm{p}}(r_{\rm{p}})$ in TNG for the same mass bins and color cut for comparison. The excess surface density distributions $\bar{n}_2w_{\rm{p}}(r_{\rm{p}})$ of red and blue galaxies around central galaxies in each mass bin, measured with PAC in both observations and TNG, are presented in Figure \ref{nwp_0_2}-\ref{nwp_5_7} for redshift ranges $0<z_s<0.2$, $0.3<z_s<0.5$, and $0.5<z_s<0.7$, respectively. In these figures, red color represents the results for red subsamples, while blue color represents the results for blue subsamples, for both observations and simulations.

In each redshift bin and mass bin, the $\bar{n}_2w_{\rm{p}}(r_{\rm{p}})$ results for red subsamples drop more sharply than blue subsamples within the projected radius $r_{\rm{p}}$ around $1 h^{-1}\rm{Mpc}$ for both observations and simulations, which means the quenched fractions decrease in this range. This shows that galaxies are more likely to be quenched if they are closer to central galaxies, which confirms the effect of environment. Furthermore, as the satellite stellar masses increase in observation, we can find a significant increase in the $\bar{n}_2w_{\rm{p}}(r_{\rm{p}})$ results of red subsamples compared to blue subsamples. This shows the monotonous increase of quenched fraction with stellar mass, which  manifest the mass quenching.

\subsection{the fraction profile}\label{red_fra}
To better illustrate the blue and red fraction distributions of galaxies around the centrals, we calculate fractions of \(\bar{n}_2w_{\rm{p}}(r_{\rm{p}})\) for red and blue subsamples separately. We employ the jackknife resampling technique mentioned earlier to estimate the statistical error of observational data. Firstly, we define the fraction of \(\bar{n}_2w_{\rm{p}}(r_{\rm{p}})\) for red or blue subsamples of the \(k\)th jackknifed realization as:
\begin{equation}
f_{\rm{k}}^{i} = \frac{\bar{n}_{2, \rm{k}}^{i} w_{\rm{p}, k}^{i}\left(r_{\rm{p}}\right)}{\bar{n}_{2, \rm{k}} w_{\rm{p}, k}\left(r_{\rm{p}}\right)}
\end{equation}

Here, \(i\) represents red or blue subsamples. Then, the mean value of the fraction of \(\bar{n}_2w_{\rm{p}}(r_{\rm{p}})\) for red or blue subsamples and the corresponding error can be expressed as:
\begin{equation}
f^{i} = \frac{1}{N_{\mathrm{sub}}} \sum_{k=1}^{N_{\mathrm{sub}}} f_{\rm{k}}^{i}\,\,,\label{eq:jack_fq}
\end{equation}
\begin{equation}
\sigma^{2}_{f^{i}} = \frac{N_{\text {sub }}-1}{N_{\text {sub }}} \sum_{k=1}^{N_{\text {sub }}}\left(f_{\rm{k}}^{i} - f^{i} \right)^{2} \,\,,\label{eq:jack_error}
\end{equation}

\begin{figure*}[ht!]
\begin{center}
\includegraphics[width=1.03\textwidth]{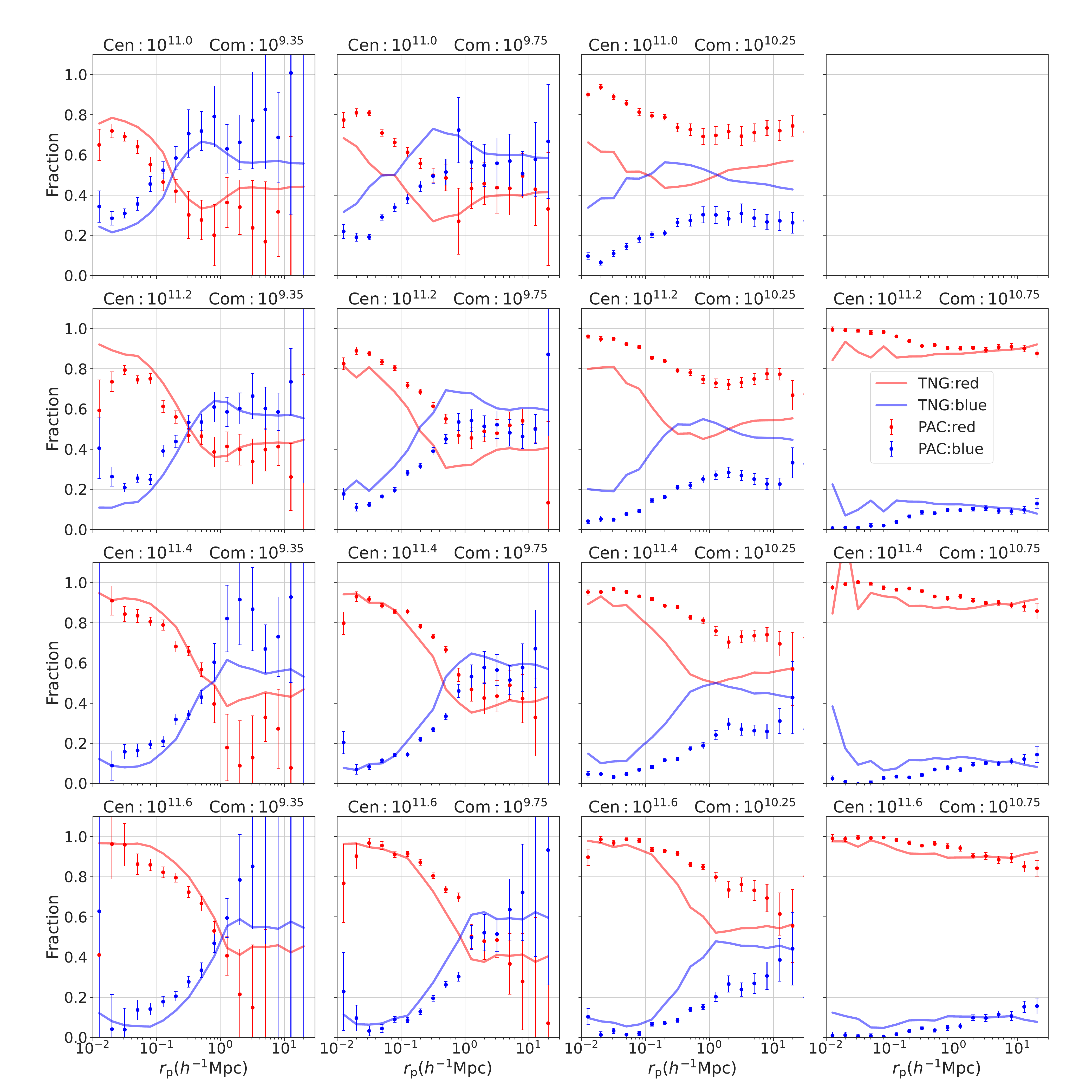}
\end{center}
\caption{The fraction of $\bar{n}_2w_{p}(r_{\rm{p}})$ for blue and red subsamples in redshift range $z_s<0.2$. The same row represents the results for the same central galaxy mass bin (from top to bottom:$[10^{10.9},10^{11.1},10^{11.3},10^{11.5},10^{11.7}]M_{\odot}$). The four columns from left to right display the results for four distinct mass bins of the photometric sample. Colored dots with error bars are the results from observations. Lines are from TNG. Here red color corresponds to red subsamples, blue color corresponds to blue subsamples for both observations and simulations.}
\label{color_nromal_0_2}
\end{figure*}

\begin{figure*}[htbp]
\begin{center}
\includegraphics[width=1.03\textwidth]{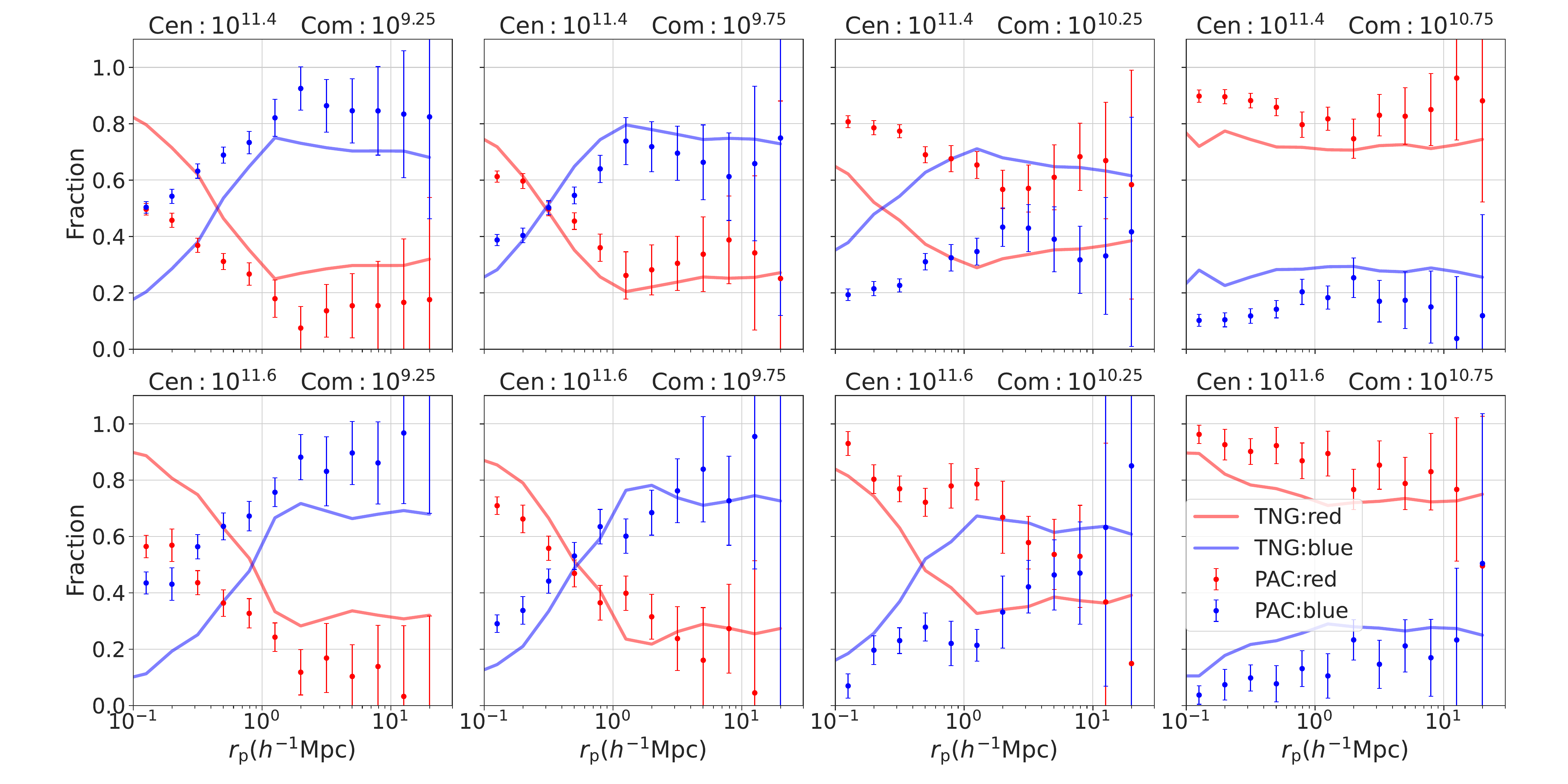}
\end{center}
\caption{The same as Figure \ref{color_nromal_0_2} but for redshift range $z_s=0.3\sim0.5$. The same row represents the results for the same central galaxy mass bin (from top to bottom: Top:$10^{11.4}M_{\odot}$, Bottom:$10^{11.6}M_{\odot}$).}
\label{color_nromal_3_5}
\end{figure*}

\begin{figure*}[htbp]
\begin{center}
\includegraphics[width=1.03\textwidth]{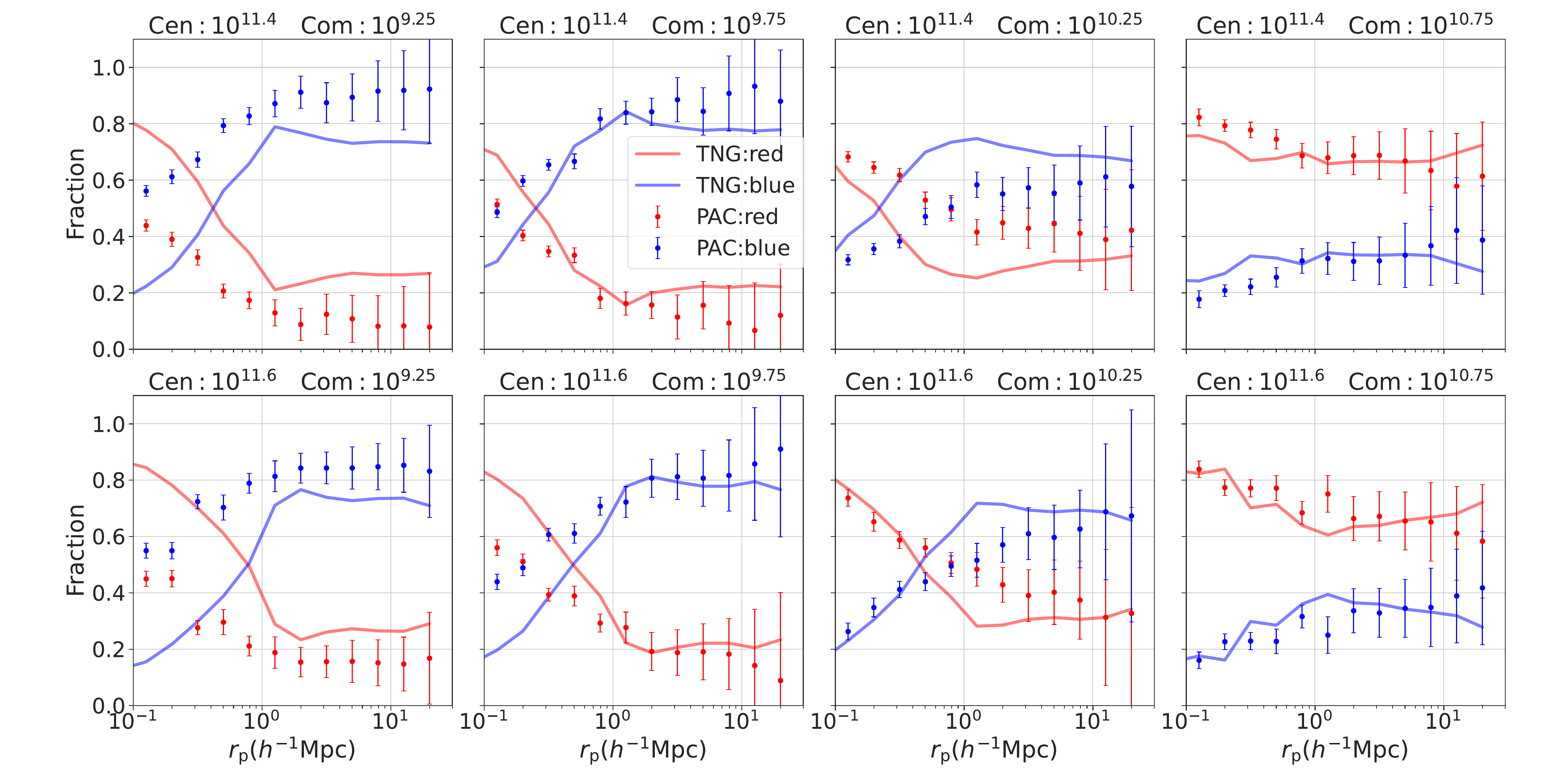}
\end{center}
\caption{The same as Figure \ref{color_nromal_3_5} but for redshift range $z_s=0.5\sim0.7$.}
\label{color_nromal_5_7}
\end{figure*}

The fraction distributions are shown in Figure \ref{color_nromal_0_2}-\ref{color_nromal_5_7}. In all these figures, the red color corresponds to the results of the red subsamples and the blue color corresponds to the blue subsamples. Dots with error bars are for the observations and lines are for TNG.

For the radial dependence, in each redshift and mass bin, it is evident that the quenched fractions of galaxies (i.e., the fractions of the red subsamples) decrease with the projected radius \(r_{\rm{p}}\) to the central galaxies for both observations and simulations. This trend reflects the influence of the environment on the star formation of galaxies. Galaxies situated closer to the massive central galaxies, i.e., closer to the center of the massive dark matter halos, are more likely to be quenched. In observations, this trend continues to around \(3r_{\rm{vir}}\). At larger distances, the quenched fractions stabilize at nearly constant levels, suggesting that the role of central galaxies (the host halos) in quenching companion galaxies is primarily within \(3r_{\rm{vir}}\). This result is consistent with \cite{2018ApJ...857...71B} who showed that environmental impact can extend to around $4.5R_{200}$. This result is very intriguing, as it is in line with the depletion radius that defines the boundary between the inner growing region of halo and the depleting environment\citep{2021MNRAS.503.4250F,2023ApJ...953...37G}.

Concerning the halo mass dependence, companion galaxies around more massive central galaxies (e.g., \(10^{11.6}M_{\odot}\)) are more likely to be quenched. This behavior becomes more evident in the total satellite fraction, as shown below. This trend is consistent with \cite{2022A&A...666A.131S} who found that the fraction of red galaxies increases with the density of the environment. Such a trend may arise from two effects: 1) because more massive halos usually possess more hot gas, ram pressure might be more effective in removing gas around satellite galaxies; and 2) more massive halos have more substructures, leading to increased harassment of companion galaxies.

Regarding the companion stellar mass dependence, as the stellar mass of companion galaxies increases, the red fraction consistently rises at all \(r_{\rm{p}}\), regardless of whether they are inside or outside the host halos of the spectroscopic central galaxies. At scales larger than \(3r_{\rm{vir}}\) of the host halos, the red fraction stabilizes at an almost constant value, representing the average value in the Universe. This pattern reflects the fact that quenching depends on stellar mass (mass quenching). In Figs \ref{color_nromal_0_2}-\ref{color_nromal_5_7}, it is evident that, although the dense halo environment always quenches an additional fraction of galaxies relative to the background mean, the slight increase in the quenched fraction for massive companion galaxies does not imply that the environment is less effective for massive galaxies. Instead, it results from the fact that, due to mass quenching, fewer star-forming galaxies are left that can be influenced by the environment, as we will demonstrate in the next subsection.

\subsection{Quenched fraction excess}\label{sec:QFE}
To better characterize galaxy quenching driven by environmental factors, we compute the quenched fraction excess (QFE) that describes the fraction of the star forming satellite galaxies quenched due to a dense environment as follows:

\begin{equation}
QFE = \frac{{f_{\rm{q}}}-\bar{f_{\rm{q}}}}{1-\bar{f_{\rm{q}}}}\,\,. \label{qe}
\end{equation}

Here, \(f_q\) represents the quenched fraction (i.e., the red fraction obtained in \ref{red_fra} for simplicity). \(\bar{f_{\rm{q}}}\) is the average quenched fraction of galaxies with the same stellar mass \(M_{*}\) at redshift \(z\). In previous studies\citep[e.g.,][]{2010ApJ...721..193P,2016MNRAS.456.4364B,2021MNRAS.506.3364R}, \(\bar{f_{\rm{q}}}\) is chosen as the fraction of quiescent galaxies in the field. However, in this paper, as mentioned in \ref{red_fra}, massive dark matter halos mainly affect the star formation of their satellites about \(3r_{\rm{vir}}\) away. Therefore, for each companion mass bin, we select three points located just beyond \(2r_{\rm{vir}}\) in each central mass bin (e.g., we select a total of six points in two rows of the first column in Figure \ref{color_nromal_5_7}) and calculate the weighted average value for all these points to represent the average quenched fraction \(\bar{f_{\rm{q}}}\). The details are as follows:

Firstly, for the companion stellar mass \(M_{*,\rm{com}}\) at redshift \(z_s\), we select the three points located just beyond \(2r_{\rm{vir}}\) in each central mass bin based on Equations \ref{eq:jack_fq} and \ref{eq:jack_error}, which can be written more explicitly as:

\begin{equation}
{f}_{\rm{q}}^{\rm{c,j}} = \frac{1}{N_{\rm{sub}}}  \sum_{k=1}^{N_{\rm{sub}}} 
f_{\rm{q}}^{\rm{c,j,k}}
\end{equation}

\begin{equation}
\sigma^{2}_{f_{\rm{q}}^{\rm{c,j}}}=\frac{N_{\mathrm{sub}}-1}{N_{\mathrm{sub}}} \sum_{k=1}^{N_{\mathrm{sub}}}\left(f_{\rm{q}}^{\rm{c,j,k}}-f_{\rm{q}}^{\rm{c,j}}\right)^2
\end{equation}

Here, \(k\) represents the \(k\)th jackknifed realization, \(j\) represents the three closest points larger than \(2r_{\rm{vir}}\), and \(c\) is the index of the central mass bins.

Then, we collect all the points in all central mass bins and get the inverse variance-weighted average value of the average quenched fraction \(\bar{f_{\rm{q}}}\) and the error:

\begin{equation}
\bar{f}_{\rm{q}} = \frac{1}{\sum_{p=1}^{C\times J}1/\sigma^2_{f_{\rm{q}}^{\rm{p}}}}{\sum_{p=1}^{C\times J}}{f_{\rm{q}}^{\rm{p}}} / \sigma^2_{f_{\rm{q}}^{\rm{p}}}
\end{equation}

\begin{equation}
\sigma^{2}_{\bar{f_{\rm{q}}}} = \frac{1}{\sum_{p=1}^{C\times J}1/\sigma^2_{f_{\rm{q}}^{\rm{p}}}}
\end{equation}

Here, \(C\) is the number of central mass bins adopted for each companion mass bin, and \(J=3\), following the definition above for the points around \(3r_{vir}\). Therefore, \(C\times J\) represents the total number of points collected for the companion stellar mass bin. 

With the quenched fraction results obtained above, we calculate the QFE for different stellar mass bins of central and companion galaxies. With the QFEs, we can quantify the efficiency that different halo environments quench galaxies of different stellar mass. Based on Equation \ref{qe}, the $\sigma_{\rm{QFE}}$ is estimated from the error propagation of the  quenched fraction $f_{\rm{q}}$ and the average quenched fraction $\bar{f_{\rm{q}}}$. We also apply the same method to the TNG simulation to get the corresponding results.

\begin{figure*}[htbp]
\begin{center}
\includegraphics[width=1.03\textwidth]{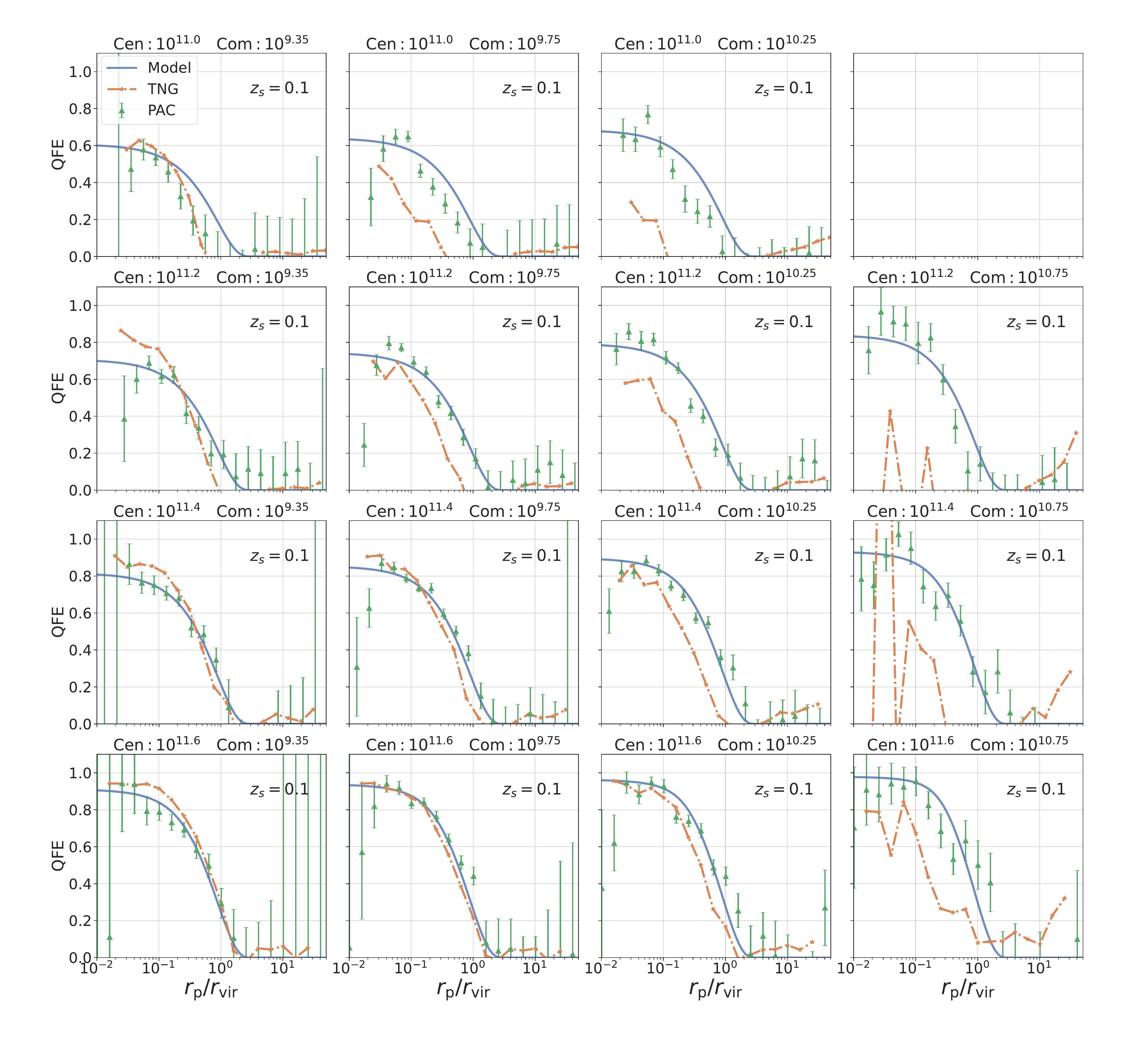}
\end{center}
\caption{The QFE results of observation(green triangles with error bar), fitting model(blue lines) and TNG(orange dotted lines) in redshift range $z_s<0.2$. The four columns from left to right show the results for four companion stellar mass bins from $10^{9.2}M_{\odot}$ to $10^{11.0}M_{\odot}$. The rows from top to bottom display the results for different central stellar mass bins from $10^{10.9}M_{\odot}$ to $10^{11.7}M_{\odot}$. Here we use the mean values to present in each panel. The title of each subplot indicates the corresponding central and companion stellar mass. The mean values of $r_{vir}$ for Jiutian and TNG are listed in Table \ref{rvir_jiutian_TNG}.}
\label{quench_ef1}
\end{figure*}

\begin{figure*}[htbp]
\begin{center}
\includegraphics[width=1.03\textwidth]{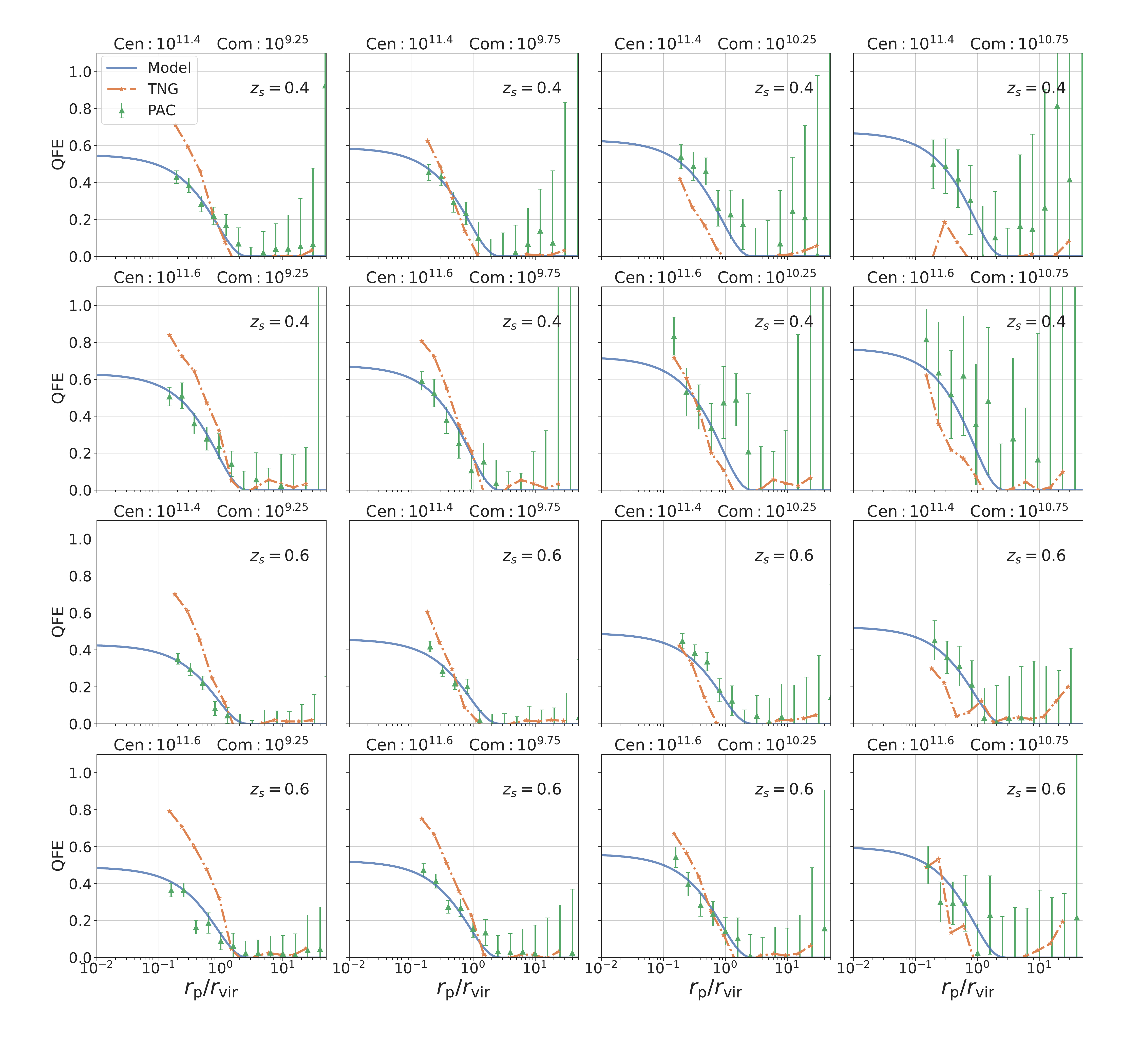}
\end{center}
\caption{The same as Figure \ref{quench_ef1} but for different redshift ranges($0.3<z_s<0.5$ and $0.5<z_s<0.7$). The four columns from left to right show the results for four companion stellar mass bins from $10^{9.0}M_{\odot}$ to $10^{11.0}M_{\odot}$. The rows from top to bottom display the results for these two redshift bins and for different central stellar mass bins from $10^{11.3}M_{\odot}$ to $10^{11.7}M_{\odot}$. The title of each subplot indicates the corresponding central and companion stellar mass.}
\label{quench_ef2}
\end{figure*}

\begin{figure*}[htbp]
\begin{center}
\includegraphics[width=0.8\textwidth]{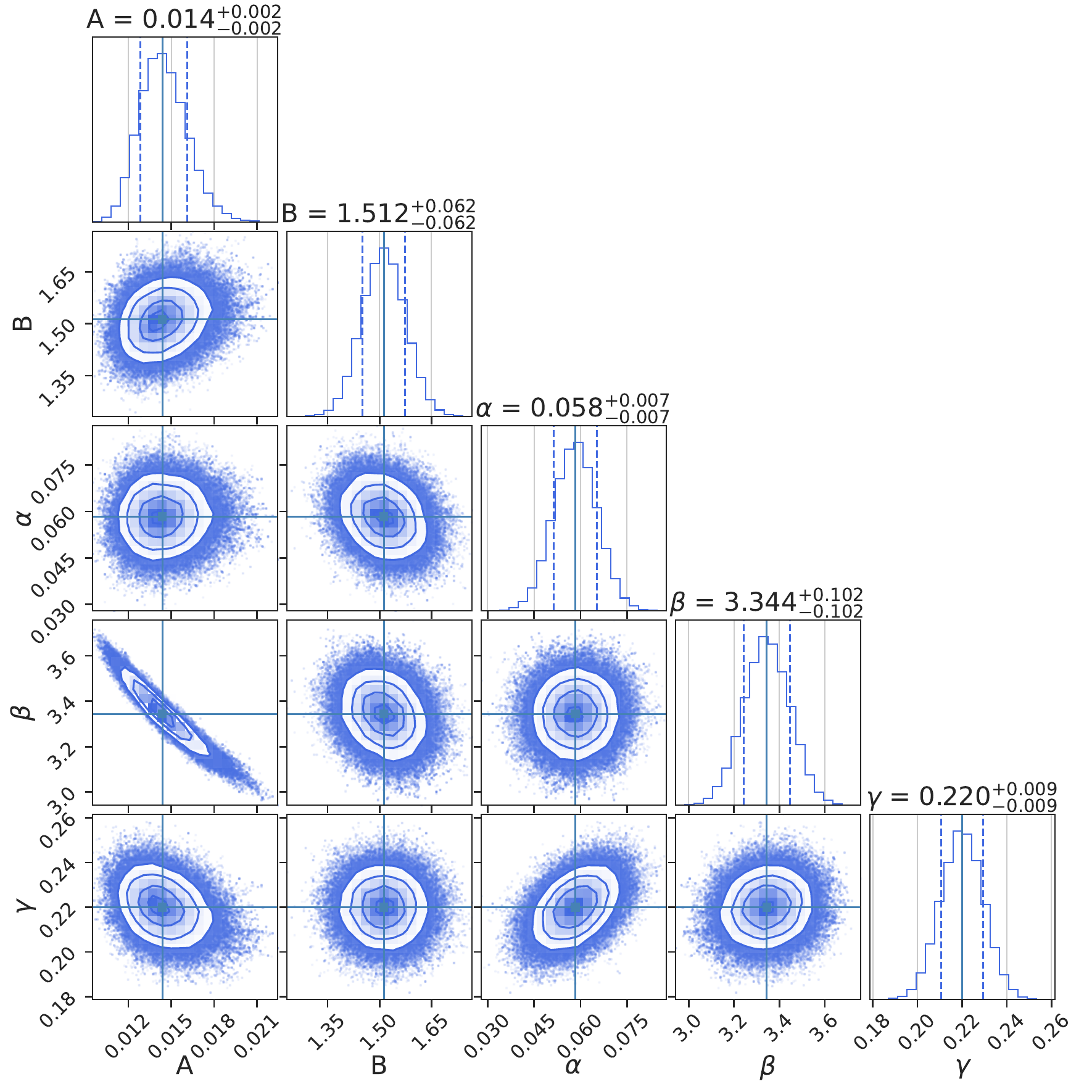}
\end{center}
\caption{The posterior distributions of the parameters in the fitting model. The 1D PDF of each parameter is plotted as a histogram at the top panel of each column, where the median value and $1\sigma$ uncertainty is also labeled.The 2D joint PDF of each parameter pair is shown as a contour with three confidence levels $68\%$, $95\%$ and $99\%$.}
\label{quench_parameter}
\end{figure*}

Figures \ref{quench_ef1} and \ref{quench_ef2} demonstrate the QFE results for different stellar mass bins of central and companion galaxies at three redshifts. For a given central mass (i.e. environment), the QFE is very similar for different satellite galaxies, though there is a rather weak trend that more massive satellites are more easily to be quenched. This trend, though rather weak, is counterintuitive, as one would expect that the less massive ones are more vulnerable to the dense environment. Although we have not fully understood the cause, we would point out that the quenching efficiency certainly depends on the definition of quenching (the line in Figure \ref{redshift_colorcut}), and it may take more time for the small ones to be quenched to across the definition criterion. For a given companion mass, we can see that the more massive halos are more effective in the quenching even after the scale is normalized by $r_{\rm vir}$. This can be understood, as we pointed before, as the ram pressure and the harassment effects are likely more effective in more massive halos. Comparing those of the same stellar bins at different redshifts, we can easily find the quenching efficiency is higher at the lower redshift.  As the SHMR changes very slowly in the redshift range of our interest, the correspond halo mass or radius does change very little (Table \ref{rvir_jiutian_TNG}). The significant change of the QFE with redshift may reflect the fact that the satellite galaxies at low redshift have a longer time to quench in the dense environment than those at higher redshift, since the halo accretion rate slows with redshift for a given halo mass effect \citep{2003ApJ...597L...9Z,2009ApJ...707..354Z}.


To summarize, the QFE relies on the halo-centric distance, redshift, mass of host halos and stellar mass of the companions. Taking into account these dependencies, we propose a fitting model for the QFE that hinges on the companion stellar mass $M_{\rm{*, com}}$, the central stellar mass $M_{\rm{*, cen}}$, redshift $z_s$, viral radius $r_{\rm{vir}}$, viral mass $M_{\rm{vir}}$ and the projected distance $r_{\rm{p}}$ as follows:

\begin{equation}\label{limit}
Q F E=\left\{\begin{array}{lr}
\left(\frac{Func^{10}}{Func^{10}+1}\right)^{0.1}, & r_{\mathrm{p}}<=3 r_{\rm{vir}} \\
0, & r_{\rm{p}}>3 r_{\rm{vir}}
\end{array}\right.
\end{equation}

where 
\begin{equation}
\begin{aligned}
&Func =  \\
& \frac{A}{(1+z_s)^{B}}\left(\frac{{M_{\rm{*, com}}}}{{M_{\rm{*,cen}}}}\right)^\alpha\times \left(\frac{M_{\rm{vir}}}{10^{12}h^{-1}M_{\odot}}\right)^\gamma\times \left(3-\frac{r_{\rm{p}}}{r_{\rm{vir}}}\right)^\beta
\nonumber
\end{aligned}
\end{equation}

Here we use the function $Func$ to describe these dependencies and we use an additional limiting function in Equation \ref{limit} to guarantee that the QFE is less than 1 for all stellar mass bins. The limiting function effectively maintains the results for small stellar mass bins and places reasonable constraints on large mass bins. It makes sense because we adopt that the satellite galaxies in the most inner region of massive host halos are almost red. Here we use $3r_{\rm{vir}}$ as the critical point according to the definition of the QFE in equation \ref{qe}.

In order to construct the fitting model, we use the Markov Chain Monte Carlo (MCMC) sampler emcee\citep{2013PASP..125..306F}. The fitting results are shown in Figure \ref{quench_ef1} and \ref{quench_ef2} as blue lines. Posterior probability density functions (PDFs) of the parameters of our model from the MCMC are shown in Figure \ref{quench_parameter} and Table \ref{parameter}. $\chi^{2} / dof = 0.939$. All of the parameters are constrained well and the fitting is almost good for all mass bins in all redshift ranges.

\begin{table*}
\caption{Posterior PDFs of the Parameters from Markov Chain Monte Carlo with their $1\sigma$ uncertainties.}\label{parameter}
\centering
\begin{tabular}{cccccc}\\
\hline\hline
\colhead{model} & \colhead{A} & \colhead{B} & \colhead{$\alpha$} & \colhead{$\beta$} & \colhead{$\gamma$}\\
\hline
 values & $0.014_{-0.002}^{+0.002}$ & $1.512_{-0.062}^{+0.062}$ & $0.058_{-0.007}^{+0.007}$ & $3.344_{-0.102}^{+0.102}$ & $0.220_{-0.009}^{+0.009}$ \\
\hline
\end{tabular}
\end{table*}

\subsection{the total satellite quenched fraction}\label{2D_fq}
We are now focusing on the total satellite quenched fraction. Satellites are defined as the galaxies within the virial radius $r_{\rm{vir}}$ of the dark matter halos of the central galaxies.

To calculate the satellite quenched fraction, we need to compute the galaxy number counts within the virial radius $r_{\rm{vir}}$ for all satellites and for quenched ones respectively.  Thus we need to drop out the galaxies beyond $r_{\rm{vir}}$ along the line-of-sight from the projected $\bar{n}_2w_{\rm{p}}(r_{\rm{p}})$. Here we use the method provided by \citet{2012ApJ...760...16J} to obtain the number of real satellites and quenched satellites within the distance $r_{\rm{vir}}$ independently. Firstly, we assume the real space correlation function $\xi(r)$ follows the power-law form $\xi(r) = (r/r_0)^{-\gamma}$, then the overall projected correlation function becomes:
\begin{equation}
    w_{\mathrm{p}}\left(r_{\mathrm{p}}\right)=r_{\mathrm{p}}\left(\frac{r_{\mathrm{p}}}{r_{0}}\right)^{-\gamma} \Gamma\left(\frac{1}{2}\right) \Gamma\left(\frac{\gamma-1}{2}\right) / \Gamma\left(\frac{\gamma}{2}\right) 
    \label{wp}
\end{equation}

By fitting the projected radial distribution profile of satellites in the range $0.1<r_{\rm{p}}<1 h^{-1}\rm{Mpc}$, we can get the correlation length $r_0$ and the power-law index $\gamma$.
Next, considering $w_{\rm{p}}$ comprises two components: one from structures within the virial radius $r_{\rm{vir}}$, the other from the correlated nearby structures outside of the virial radius along the line of sight, $w_{\rm{p}}$ could be written as:
\begin{equation}
    w_{\mathrm{p}}\left(r_{\mathrm{p}}\right)=w_{\mathrm{p} 1}\left(r_{\mathrm{p}}\right)+w_{\mathrm{p} 2}\left(r_{\mathrm{p}}\right)
\end{equation}
where
\begin{equation}\label{wp1}
    w_{\mathrm{p} 1}\left(r_{\mathrm{p}}\right)=2 \int_{r_{\mathrm{p}}}^{r_{\mathrm{vir}}} r d r \xi(r)\left(r^{2}-r_{\mathrm{p}}^{2}\right)^{-1 / 2}
\end{equation}
and 
\begin{equation}
    w_{\mathrm{p} 2}\left(r_{\mathrm{p}}\right)=2 \int_{r_{\mathrm{vir}}}^{\infty} r d r \xi(r)\left(r^{2}-r_{\mathrm{p}}^{2}\right)^{-1 / 2}
\end{equation}

Through taking the fitting $\xi(r) = (r/r_0)^{-\gamma}$ into Equation \ref{wp1}, we can get $w_{\rm{p}1}$ and the ratio $w_{\rm{p}1}(r_{\rm{p}})/w_{\rm{p}}(r_{\rm{p}})$. Afterwards, the total number of real satellites within the projected distance $r_{\rm{vir}}$ can be obtained:
\begin{equation}
    N\left(<r_{\mathrm{vir}}\right)=\int_{0}^{r_{\mathrm{vir}}} \Sigma\left(r_{\mathrm{p}}\right) \frac{w_{\mathrm{p} 1}\left(r_{\mathrm{p}}\right)}{w_{\mathrm{p}}\left(r_{\mathrm{p}}\right)} 2 \pi r_{\mathrm{p}} d r_{\mathrm{p}}
\end{equation}
By applying above steps on $\bar{n}_2w_{\rm{p}}(r_{\rm{p}})$ of all samples and the red subsamples separately, we obtain the number of real satellites and quenched satellites accordingly. Finally, we can calculate the quenched fractions for observation:
\begin{equation}
    F_q = \frac{N_{\rm{red}}\left(<r_{\mathrm{vir}}\right)}{N_{\rm{all}}\left(<r_{\mathrm{vir}}\right)}
\end{equation}
We also apply this method on TNG for comparison. Because we obtain the fractions from the surface distribution $\bar{n}_2w_{\rm{p}}(r_{\rm{p}})$, this method could be understood as a 2D method. 

To verify the validity of the 2D method, we calculate the satellite quenched fractions in TNG directly and confirm the reliability of it. The comparison result are shown in APPENDIX \ref{TNG_2d_3d}.

Figure \ref{Fq} shows the satellite quenched fraction for different mass bins and for three different redshift bins. Each panel shows the quenched fractions for the same redshift bin. The results for observation are represented by dots, those for TNG using the 2D method are represented by solid lines. Different colors represent different central stellar mass bins.


\begin{figure*}[htbp]
\begin{center}
\includegraphics[width=1.03\textwidth]{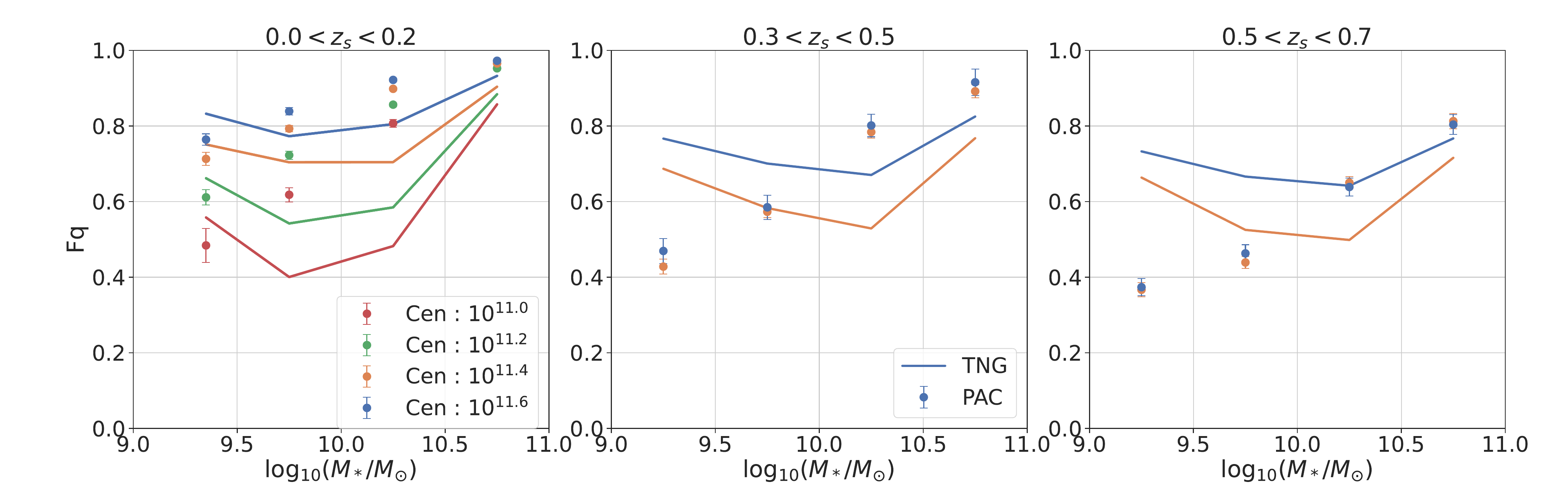}
\end{center}
\caption{The satellite quenched fractions of observation and TNG for three distinct redshift bins: $0<z_s<0.2$(left), $0.3<z_s<0.5$(medium), and $0.5<z_s<0.7$(right). Colored dots with errorbars are the results from observations. Solid-lines are for TNG 2D method. For both observations and simulations, different colors are used for the corresponding central stellar mass bins: red($10^{11.0}M_{\odot}$), green($10^{11.2}M_{\odot}$), orange($10^{11.4}M_{\odot}$) and blue($10^{11.6}M_{\odot}$).}
\label{Fq}
\end{figure*}

In observation, the dependence of quenched fraction on both central and satellite masses is consistent with what we mentioned above. The satellite quenched fractions increase monotonously with satellite mass. The quenched fractions also increase with the stellar mass of central galaxies. We also find a redshift evolution of quenched fractions, indicating that galaxies at lower redshift become redder and have higher quenched fractions. 

\section{Comparing to TNG300-1}\label{sec:Discussion}
To comprehensively assess our observational results, we conduct a detailed comparison with the Illustris TNG300-1 simulation. In this section, we present the key findings of this comparison. 

Firstly, examining the excess surface density distribution, $\bar{n}_2w_{\rm{p}}(r_{\rm{p}})$, and the fraction distributions, we observe a clear redshift dependence in both observations and simulations. The red fractions consistently increase with decreasing redshift. Regarding the radial dependence, the red fractions in TNG decrease more rapidly with $r_{\rm{p}}$ than in observations, indicating a more pronounced increase toward the centers. This may arise from premature stripping of gas around companion galaxies, as also found by \cite{2020MNRAS.495.4570M}. Concerning the mass dependence, in observations, as companion photometric stellar masses increase, companion galaxies around massive central galaxies exhibit larger red fractions at all $r_{\rm{p}}$. This trend persists from $10^{9.0}M_{\odot}$ to $10^{11.0}M_{\odot}$. However, in TNG, this trend is not consistently observed, with the red fractions of  less massive companion galaxies ($10^{9.0}\sim10^{9.5}M_{\odot}$) sometimes exceeding those of the more massive ones ($10^{10.0}\sim10^{10.5}M_{\odot}$).

Secondly, examining the QFE results, we find that both observations and simulations exhibit a clear redshift dependence, with the QFE increasing as redshift decreases. Regarding the influence of the environment, the QFE is notably affected by the central stellar mass in all three redshift ranges, rising with increasing mass in both observations and simulations. This indicates that more massive halos are more effective in quenching. Regarding the dependence on companion stellar masses, the QFE results show a weak trend in observations, where more massive companion galaxies are more likely to be quenched from $10^{9.0}\sim10^{11.0}M_{\odot}$. Conversely, the QFE results in TNG exhibit an entirely opposite trend, with the QFE dropping significantly as companion stellar masses increase.

Next, examining the total quenched fraction of satellites, $F_{\rm{q}}$, in both observations and simulations, we observe a redshift evolution where satellite galaxies become redder and exhibit higher quenched fractions at lower redshifts. Regarding the dependence on mass, $F_{\rm{q}}$ increases with central stellar mass, and this dependence is more pronounced in TNG than in observations. Concerning satellite stellar mass, $F_{\rm{q}}$ monotonously increases in observations, whereas this is not consistently observed in TNG. This suggests that environmental effects do not alter the mass dependence of galaxy color distribution in observations, while environmental effects in TNG appear to be too strong on small satellites.

\section{Conclusion} \label{sec:Coclusion}
In this paper, we use the PAC method to calculate the excess surface density $\bar{n}_2w_{\rm{p}}(r_{\rm{p}})$ of companion galaxies with different mass and colors around massive central galaxies from redshift $z_s=0.7$ to the present time. Combining these results, we get the corresponding quenched fractions. Moreover, based on the quenched fraction results, we obtain the QFE and construct a fitting model. Finally, we calculate the satellite quenched fractions for different redshift and mass bins. The results are compared with the Illustris TNG300-1 simulation. Our main results are listed below:
\begin{itemize}
    \item Both mass and environment have effects on the quenching of the companion galaxies in observations and simulations. In order to separate them, we use the QFE to quantify the environment quenching.
    \item We find that the high density host halo environment takes effect in quenching the star formation of the companions up to the scale about $3r_{\rm{vir}}$. We also provide a fitting model of the QFE which can approximately describe the dependencies on the halo-centric distance, redshift, host halo mass and the companion stellar mass.
    \item In observation the quenched fractions $F_q$ of satellites increase monotonously with the satellite stellar mass, but this trend is not found in TNG, indicating too strong an environment effect on small satellites in TNG.
    
\end{itemize}
In conclusion, mass and environment quenching can both affect the evolution of companion galaxies. Environment quenching is a wide-ranging effects that can affect the star formation of galaxies up to $3r_{\rm{vir}}$ from the halo center. Results from the IllustrisTNG simulation are deviated from these observational results especially for small companions, which may indicate that the galaxy quenching recipes and/or simulation resolutions should be improve.  Our fitting model can help to test and improve the current galaxy formation models.

In the future, with the deeper and wider spectroscopic and photometric surveys, we can extend our studies to fainter galaxies and higher redshift, taking advantage of PAC.

\section*{Acknowledgement}
The work is supported by NSFC (12133006, 11890691), by National Key R\&D Program of China (2023YFA1607800, 2023YFA1607801), and by 111 project No. B20019. This work made use of the Gravity Supercomputer at the Department of Astronomy, Shanghai Jiao Tong University. We gratefully acknowledge the support of the Key Laboratory for Particle Physics, Astrophysics and Cosmology, Ministry of Education.We acknowledge the science research grants from the China Manned Space Project with NO. CMS-CSST-2021-A03. We are grateful for useful discussions on galaxy catalog with Wenting Wang. Z.Y. thanks Xiaokai Chen and Jian Yao for kind help.

The Hyper Suprime-Cam (HSC) collaboration includes the astronomical communities of Japan and Taiwan, and Princeton University. The HSC instrumentation and software were developed by the National Astronomical Observatory of Japan (NAOJ), the Kavli Institute for the Physics and Mathematics of the Universe (Kavli IPMU), the University of Tokyo, the High Energy Accelerator Research Organization (KEK), the Academia Sinica Institute for Astronomy and Astrophysics in Taiwan (ASIAA), and Princeton University. Funding was contributed by the FIRST program from Japanese Cabinet Office, the Ministry of Education, Culture, Sports, Science and Technology (MEXT), the Japan Society for the Promotion of Science (JSPS), Japan Science and Technology Agency (JST), the Toray Science Foundation, NAOJ, Kavli IPMU, KEK, ASIAA, and Princeton University.

This publication has made use of data products from the Sloan Digital Sky Survey (SDSS). Funding for SDSS and SDSS-II has been provided by the Alfred P. Sloan Foundation, the Participating Institutions, the National Science Foundation, the U.S. Department of Energy, the National Aeronautics and Space Administration, the Japanese Monbukagakusho, the Max Planck Society, and the Higher Education Funding Council for England.

\appendix
\section{result comparison}\label{nwpcheck}
We calculate $\bar{n}_2w_{\rm{p}}(r_{\rm{p}})$ using the spectroscopic data directly to compare with the PAC result here. We use the Main sample mentioned before and choose those galaxies located in the continuous region of the NGC. Considering completeness of the Main sample, we select two satellite samples: galaxies with stellar mass $10^{10.0} M_{\odot}< M_{\star}<10^{10.5} M_{\odot}$ at $z_s<0.075$ and galaxies with stellar mass $10^{10.5} M_{\odot}< M_{\star}<10^{11.0} M_{\odot}$ at $z_s<0.085$. We select galaxies with stellar mass  $10^{11.3} M_{\odot}< M_{\star}<10^{11.5} M_{\odot}$ as the central sample. Then we obtain $\bar{n}_2w_{\rm{p}}(r_{\rm{p}})$ by calculating the mean number density $\bar{n}_2$ and the projected cross-correlation function $w_{\rm{p}}(r_{\rm{p}})$ separately. After that, we compare the $\bar{n}_2w_{\rm{p}}(r_{\rm{p}})$ calculated using PAC in the same stellar mass bin at $z_s = 0.1$. The comparison results are respectively shown in Figure \ref{pac_comparision_075} and \ref{pac_comparision_085}. PAC measurements are shown in lines and direct spectroscopic measurements are shown in dots. The misalignment for blue galaxies samples in Figure \ref{pac_comparision_085} may be due to the relatively small number of blue galaxies. Except this, the other results fit remarkably well for all red, blue and full samples, which gives a direct test of PAC and confirms its practicability. 

\renewcommand\thefigure{\Alph{section}\arabic{figure}}
\setcounter{figure}{0}
\begin{figure*}[htbp]
\begin{center}
\includegraphics[width=0.8\textwidth]{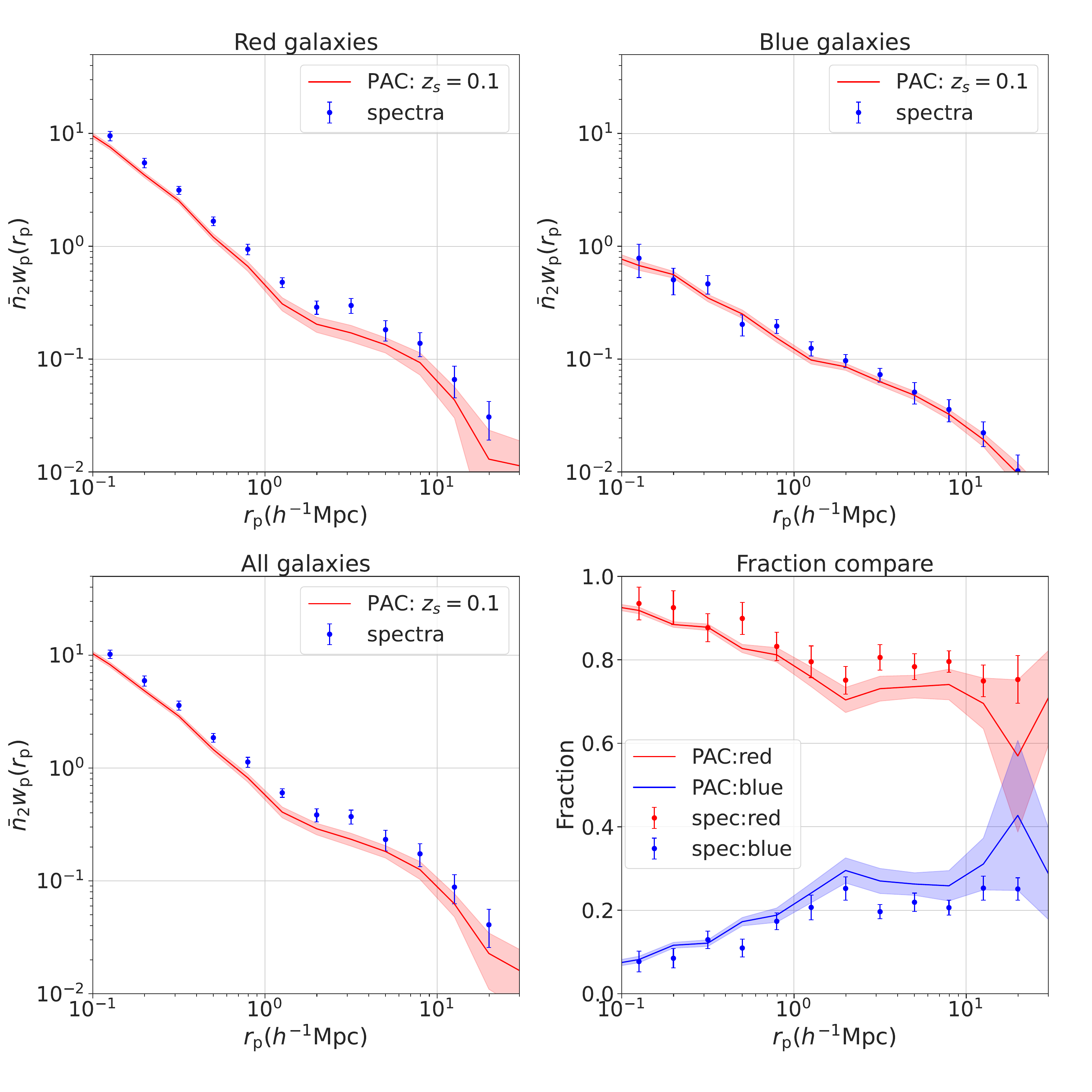}
\end{center}
\caption{Comparison of $\bar{n}_2w_{\rm{p}}(r_{\rm{p}})$ estimated from PAC and the direct calculation from the spectroscopic sample at $z_s<0.075$, with central stellar mass range $10^{11.3} M_{\odot}< M_{\star}<10^{11.5} M_{\odot}$ and satellite stellar mass range $10^{10.0} M_{\odot}< M_{\star}<10^{10.5} M_{\odot}$. The two panel in top row show results for red and blue subsamples respectively. The bottom left panel displays the result for the whole sample. In these three panels, data points represent the results by using spectroscopic directly. Lines with shadow are the results from PAC. The bottom right panel shows the normalized fraction for red and blue subsamples separately. In this panel, red line corresponds to PAC result for red galaxies, blue line corresponds to PAC result for blue galaxies, dots with error bar are for spectroscopic results.}
\label{pac_comparision_075}
\end{figure*}

\setcounter{figure}{1}
\begin{figure*}[htbp]
\begin{center}
\includegraphics[width=0.8\textwidth]{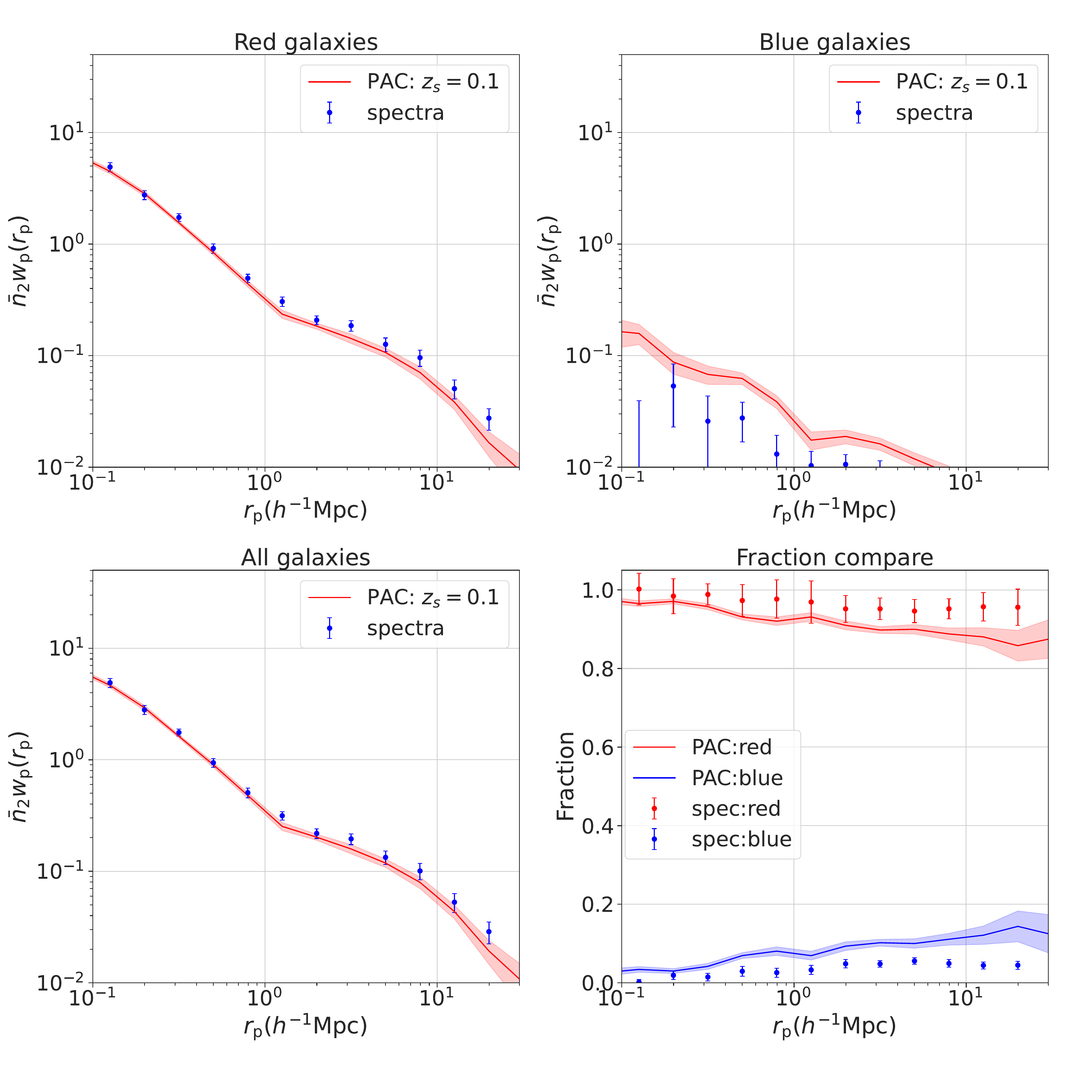}
\end{center}
\caption{The same as Figure \ref{pac_comparision_075} but for the spectroscopic sample at $z_s<0.085$, with central stellar mass range $10^{11.3} M_{\odot}< M_{\star}<10^{11.5} M_{\odot}$ and satellite stellar mass range $10^{10.5} M_{\odot}< M_{\star}<10^{11.0} M_{\odot}$.}
\label{pac_comparision_085}
\end{figure*}

\section{TNG satellite quenched fraction}\label{TNG_2d_3d}
As mentioned in Section \ref{2D_fq}, to verify the validity of the 2D method, we calculate the satellite quenched fractions in TNG directly. Firstly, we choose the central galaxies and their satellite galaxies in the needed mass bins. Then we select the satellites within $r_{\rm{vir}}$ of the central galaxies and get the number of total satellites. After that, we apply Equation \ref{colorcut_eq} to  the satellite samples to obtain the quenched satellites. Finally, the quenched fractions can be obtained through dividing the number of red satellites by the total number of satellites. We refer to this method as a 3D method because we are using spatial positions here. The comparison result are shown in Figure \ref{TNG_2d_3d_picture}, it can be seen that the two results for simulation are essentially identical. The relative errors for different mass bins are less than $6.2\%$, which confirms the correctness of the 2D method.

\renewcommand\thefigure{\Alph{section}\arabic{figure}}
\setcounter{figure}{0}
\begin{figure*}[htbp]
\begin{center}
\includegraphics[width=1.1\textwidth]{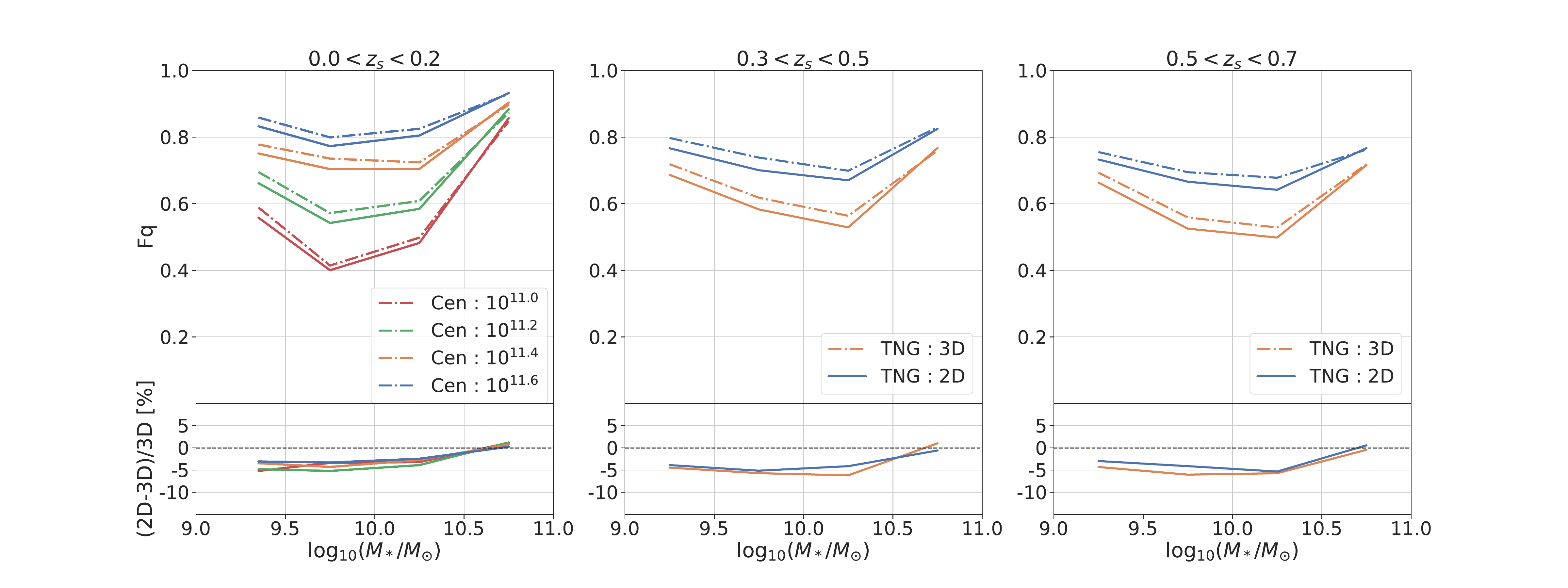}
\end{center}
\caption{The satellite quenched fraction of TNG for three redshift bins:$0<z_s<0.2$(left), $0.3<z_s<0.5$(medium), and $0.5<z_s<0.7$(right). Solid-lines are for TNG 2D method, dotted-lines are for 3D method. Different colors are used for the corresponding central stellar mass bins: red($10^{11.0}M_{\odot}$), green($10^{11.2}M_{\odot}$), orange($10^{11.4}M_{\odot}$) and blue($10^{11.6}M_{\odot}$).}
\label{TNG_2d_3d_picture}
\end{figure*}

\bibliography{sample631}{}
\bibliographystyle{aasjournal}



\end{document}